# ACCELERATED MICROPOLAR FLUID--FLOW PAST AN UNIFORMLY ROTATING CIRCULAR CYLINDER

**Abuzar Abid Siddiqui**
*Department of Basic Sciences, College of Engg. & Tech., B. Z. University, Multan-60800, Pakistan.*

**Abstract**
*In this paper, we formulated the non-steady flow due to the uniformly accelerated and rotating circular cylinder from rest in a stationary, viscous, incompressible and micropolar fluid. This flow problem is examined numerically by adopting a special scheme comprising the Adams-Bashforth Temporal Fourier Series method and the Runge-Kutta Temporal Special Finite-Difference method. This numerical scheme transforms the governing equation for micropolar fluids for this problem into system of finite-difference equations. This system was further solved numerically by point SOR-method. These results were also further extrapolated by the Richardson extrapolation method. This scheme is valid for all values of the flow and fluid-parameters and for all time. Moreover the boundary conditions of the vorticity and the spin at points far from the cylinder are being imposed and encountered too. The results are compared with existing results (for non-rotating circular cylinder in Newtonian fluids). The comparison is good. The enhancement of lift and reduction in drag was observed if the micropolarity effects are intensified. Same is happened if the rotation of a cylinder increases. Furthermore, the vortex-pair in the wake is delayed to successively higher times as rotation parameter α increases. In addition, the rotation helps not only in dissolving vortices adjacent to the cylinder and adverse pressure region but also in dissolving the boundary layer separation. Furthermore, the rotation reduces the micropolar spin boundary layer also.*

## *1. Introduction*

The problem of the fluid flow due to an infinite circular cylinder which is uniformly accelerated from rest is classical one. In 1908, [1] (for the first time) considered the general problem for two cases of motion from rest. In case 1, he studied the flow past the impulsively started circular cylinder with uniform velocity while in case 2, he investigated the impulsively started circular cylinder with uniform acceleration. For the second case [2–3] showed also their efforts numerically. Görlter [4–5] generalised this theory to other types of variation of the initial velocity of the cylinder. The numerical methods used to solve Navier-Stokes equations for these types of flows were very approximate and valid for all Reynolds number only for leading term in the series solution, which is expanded in the powers of time. But subsequent terms in the expansion are valid only for infinite Reynolds number.

In 1974, [6] has studied the second case of [1] very well and improved and solved them by two techniques. In the first, spectral method while in the second direct integration of Navier-Stokes equations which was the extention of first method. Experimentally this problem was also examined by [7] in 1972. He measured time of separation of flow and the growth of the separated wake for Reynolds numbers $R^2$=97.5, 5850 and 122x10$^3$.

The problem of the fluid-flow due to uniformly accelerated and rotating an infinite circular in the stationary incompressible, viscous fluids is of fundamental interest owing to its valuable and large number of applications: For example, (i) in enhancing the lift [8] (i.e., in the Flettner rotor ship where rotating vertical cylinders were employed to develop a thrust normal to any winds blowing past the ship); (ii) in controlling the boundary layer separation [9–12]; (iii) in applying the "breakaway" phenomenon [13–15] on the use of rotation; (iv) in controlling the boundary layer devices such as on the flaps of V/STOL aircraft [16] and upstream of ship rudders and to overcome separation effects in a subsonic diffuser. In this attempt we study the said problem for the micropolar fluids, whose governing and constitutive equations were developed by [17]. These are the subclass of the microfluids [18] having six degree of freedom---three more than that in the Newtonian fluids. The micropolar fluids are exemplified by blood [19-20], polymers and polymeric suspension [21], rigid-rod epoxies [22], colloidal suspension and liquid crystals [23]. These fluids are being used on lab-on-a-chip [24–25] nowadays. The significant contributions in the field of micropolar-fluid-dynamics are [17], [20] and [26–36]. On motivating the significant applications of flow past the accelerated and rotating cylinder, we embarked to probe the micropolarity-effects. In addition the rotation-effects were captured owing to the microspin of the aciculate colloids. For this, we formulated the initio-boundary-value-problem (IBVP) and solved it by the numerical scheme comprising two high order aforementioned methods. Moreover, we examined the effects on surface forces, changes in flow behaviour, variation of secondary vortices, and pressure effects especially in the boundary layer for micropolar fluids. In addition it was studied the spin-behaviour and its variation with respect to Reynolds number R, rotation parameter $\alpha$ and time t. We also probed what will be the flow behaviour and its related properties before and after the imposing of rotation of a circular cylinder. Moreover we examined; what will be the deformation, validity, and stability of numerical scheme for micropolar fluids? It was observed that, the problem under consideration is so sensitive and complicated in the sense that most the numerical schemes do not support well for all time t and Reynolds number R. But the numerical scheme, which we have adopted, has the advantage that it converges for all values of *R*, *t* and *α*.

Although the calculations were made for various values of Reynolds number *R* in the range $1 \leq R \leq 350$ and several values of *α* in the range $0 \leq \alpha \leq 4.5$, yet we presented the results for four values of $R \in \{1, 10, 100, 349.285\}$ and three values of $\alpha \in \{2.5, 3, 4.5\}$ for discussion purposes. The results are presented in graphical as well as in tabular forms in section 8, in which we have examined the streamlines behaviour, equi-spin lines, equi-vorticity lines, vorticity variation with time, the variation of surface forces with *t* for different values of *R* and *α*, specially the vortex shedding over the surface cylinder and pressure development in the wake of a cylinder. The formulation and basic analysis of the problem under consideration is given in section 2. In section 3 we deform the boundary value problem into modified polar co-ordinates system. The numerical scheme is discussed in section 4. The details regarding the calculations and evaluation of boundary conditions on vorticity and spin far from the cylinder are given in section 5 and 6 respectively, while the brief description of numerical experiment is presented in section 7.

## 2. *Basic Analysis and Formulation*

The mechanics of the problem under consideration can briefly be stated as follows: the flow is normal to an infinite circular cylinder which is uniformly accelerated and rotating from rest in an infinite stationary viscous incompressible fluid, such that the circular cylinder is rotating with angular speed $\Omega$ while advancing from right to left with uniform acceleration $f$. The rotation and translation are started at the same time. Obviously, a cylindrical co-ordinate system will be used, where the reference frame is fixed on the cylinder such that the origin coincides with the cylinder's centre. However, the coordinate system will be modified further by taking $s=lnr$. Moreover, the flow will be considered as non-steady, two-dimensional and laminar for all time.

The continuity and the governing equations for incompressible micropolar fluid in dimensional form, in the absence of body force are given as follows:

$$\nabla^*.\mathbf{V}^* = 0 \qquad (2.1)$$

$$-(\mu+\chi)\nabla^* \mathrm{x} \nabla^* \mathrm{x} \boldsymbol{\omega}^* + \chi \nabla^* \mathrm{x} \nabla^* \mathrm{x} \boldsymbol{\nu}^* = \rho[\frac{\partial \boldsymbol{\omega}^*}{\partial t} - \nabla^* \mathrm{x}(\mathbf{V}\mathrm{x}\boldsymbol{\omega}^*)] \qquad (2.2)$$

$$(\alpha+\beta+\gamma)\nabla^*(\nabla^*.\boldsymbol{\nu}^*) - \gamma\nabla^* \mathrm{x}\nabla^* \mathrm{x}\boldsymbol{\nu}^* + \chi\boldsymbol{\omega}^* - 2\chi\boldsymbol{\nu}^* = \rho j[\frac{\partial \boldsymbol{\nu}^*}{\partial t^*} + (\mathbf{V}^*.\nabla^*)\boldsymbol{\nu}^*] \qquad (2.3)$$

where $\boldsymbol{\omega}^*$=curl$\mathbf{V}^*$, and $\mathbf{V}^*$ is the dimensional fluid velocity vector, $\boldsymbol{\nu}$ the micro-rotational vector(or spin), $\rho$ the density, j the micro-inertia (gyration parameter), $\alpha$, $\beta$, $\gamma$, $\lambda$, $\mu$, and $\chi$ are the material constants (or viscosity coefficients).

Now the above equations (2.1) – (2.3) would take the form in dimensionless form as follows:

$$\nabla.\mathbf{V} = 0 \qquad (2.4)$$

$$-\nabla \mathrm{x}\nabla \mathrm{x}\boldsymbol{\omega} + \frac{\chi}{(\mu+\chi)}\nabla \mathrm{x}\nabla \mathrm{x}\boldsymbol{\nu} = \frac{\rho c\sqrt{fc}}{(\mu+\chi)}[\frac{\partial \boldsymbol{\omega}}{\partial t} - \nabla \mathrm{x}(V\mathrm{x}\boldsymbol{\omega})] \qquad (2.5)$$

$$(\alpha+\beta+\gamma)\nabla(\nabla.\boldsymbol{\nu}) - \gamma\nabla \mathrm{x}\nabla \mathrm{x}\boldsymbol{\nu} + \chi c^2\boldsymbol{\omega} - 2\chi c^2\boldsymbol{\nu} = \rho jc\sqrt{fc}[\frac{\partial \boldsymbol{\nu}}{\partial t} + (\mathbf{V}.\nabla)\boldsymbol{\nu}] \qquad (2.6)$$

where the relation between dimensional and dimensionless variables is

$$V^* = \sqrt{fc}V, \quad x_i^* = cx_i, \quad \upsilon^* = \sqrt{\frac{f}{c}}\upsilon, \quad \omega^* = \sqrt{\frac{f}{c}}\omega, \quad t^* = \sqrt{\frac{c}{f}}t$$

and * signifies the dimensional variables while $f$ and $c$ are acceleration and radius of a circular cylinder respectively as characteristic parameters. Under the above assumptions, equations (2.4) – (2.6) in component form in cylindrical polar co-ordinates system would be,

$$\frac{1}{r}[\frac{\partial (ru_r)}{\partial r} + \frac{\partial v_\theta}{\partial \theta}] = 0 \qquad (2.7)$$

$$[\frac{\partial^2 E}{\partial r^2} + \frac{1}{r}\frac{\partial E}{\partial r} + \frac{1}{r^2}\frac{\partial^2 E}{\partial \theta^2}] - \frac{\chi}{(\mu+\chi)}[\frac{1}{r}\frac{\partial}{\partial r}(r\frac{\partial \xi}{\partial r}) + \frac{1}{r^2}\frac{\partial^2 \xi}{\partial \theta^2}]$$

$$= \frac{\rho c\sqrt{fc}}{(\mu+\chi)}[\frac{\partial E}{\partial t} + u_r\frac{\partial E}{\partial r} + \frac{v_\theta}{r}\frac{\partial E}{\partial \theta}] \qquad (2.8)$$

$$\frac{\gamma}{r}[\frac{\partial}{\partial r}(r\frac{\partial \xi}{\partial r}) + \frac{1}{r}\frac{\partial^2 \xi}{\partial \theta^2}] + \chi c^2 E - 2\chi c^2 \xi$$

$$= \rho jc\sqrt{fc}[\frac{\partial \xi}{\partial t} + u_r\frac{\partial \xi}{\partial r} + \frac{v_\theta}{r}\frac{\partial \xi}{\partial \theta}] \qquad (2.9)$$

where $u_r$, $v_\theta$ are the radial and transverse component of velocity vector respectively while $\xi$, and E are the axial components of micro-rotation (spin) and vorticity vectors respectively.

In terms of stream function $\psi$ and vorticity equations (2.7) – (2.9) would take the form as,

$$E = -\nabla^2 \psi \tag{2.10}$$

$$\nabla^2 E - \frac{c_1}{2}\nabla^2 \eta = \frac{R}{2}[\frac{\partial E}{\partial t} + \frac{1}{r}\{\frac{\partial \psi}{\partial \theta}\frac{\partial E}{\partial r} - \frac{\partial \psi}{\partial r}\frac{\partial E}{\partial \theta}\}] \tag{2.11}$$

$$\nabla^2 \eta + 2c_2(E - \eta) = c_3[\frac{\partial \eta}{\partial t} + \frac{1}{r}\{\frac{\partial \psi}{\partial \theta}\frac{\partial \eta}{\partial r} - \frac{\partial \psi}{\partial r}\frac{\partial \eta}{\partial \theta}\}] \tag{2.12}$$

where $\eta = 2\xi$, $\quad \nabla^2 = \frac{\partial^2}{\partial r^2} + \frac{1}{r}\frac{\partial}{\partial r} + \frac{1}{r^2}\frac{\partial^2}{\partial \theta^2}$

$$C_1 = \frac{\chi}{\mu + \chi}, \qquad C_2 = \frac{\chi c^2}{\gamma}$$

$$C_3 = \frac{\rho j c \sqrt{fc}}{\gamma}, \quad \text{and} \quad R = \frac{2\rho c \sqrt{fc}}{\mu + \chi}$$

where $C_1$, $C_2$, and $C_3$ are dimensionless field constants while R is the Reynolds number.

Note that if we set $\chi=\eta=0$ then the above equations would transform into the governing equations of Newtonian fluids. The boundary conditions for this problem would be

$$\psi = 0, \quad \frac{\partial \psi}{\partial r} = -\alpha, \eta = 0, \quad E_w = -\frac{6\psi_1 + H^2 E_1}{2(H^2 + H^3)} + \frac{3\alpha}{2H + H^2} \quad \text{at } r = 1$$

$$\frac{\partial \psi}{\partial r} \to t\sin\theta, \quad \frac{1}{r}\frac{\partial \psi}{\partial \theta} \to t\cos\theta, \quad E \to 0, \quad \eta = 0 \quad \text{as } r \to \infty \tag{2.13}$$

### 3. BVP in Modified Polar Coordinates System

On transforming the equations into modified polar coordinates system, let us take, $s = \ln r$, then equations (2.10) – (2.12) would deform as

$$E = -e^{-2s}\Theta^2 \psi \tag{3.1}$$

$$2\Theta^2 E - C_1 \Theta^2 \eta = \frac{R}{2}[e^{2s}\frac{\partial E}{\partial t} + \frac{\partial \psi}{\partial \theta}\frac{\partial E}{\partial s} - \frac{\partial \psi}{\partial s}\frac{\partial E}{\partial \theta}] \tag{3.2}$$

$$\Theta^2 \eta - 2e^{2s}C_2(E - \eta) = C_3[e^{2s}\frac{\partial \eta}{\partial t} + \frac{\partial \psi}{\partial \theta}\frac{\partial \eta}{\partial s} - \frac{\partial \psi}{\partial s}\frac{\partial \eta}{\partial \theta}] \tag{3.3}$$

where $\quad \Theta^2 \equiv \frac{\partial^2}{\partial s^2} + \frac{\partial^2}{\partial \theta^2}$

Obviously the boundary conditions given in (2.10) would deform as,

$$\psi = \eta = 0, \quad \frac{\partial \psi}{\partial s} = -\alpha, \quad E_w = -\frac{6\psi_1 + H^2 E_1}{2(H^2 + H^3)} + \frac{3\alpha}{2H + H^2} \quad \text{at } s = 0 \tag{3.4}$$

$$e^{-s}\frac{\partial \psi}{\partial s} \to t\sin\theta, \quad e^{-s}\frac{\partial \psi}{\partial \theta} \to t\cos\theta, \quad E \to 0, \quad \eta \to 0 \quad \text{as } s \to \infty \tag{3.5}$$

where s=0 represents the surface of the cylinder. $E_w$ is for vorticity on the surface of the cylinder while the subscript 1 denotes the point one cell away to the surface of a cylinder.

Earlier numerical attempt [6] was related to the problem under consideration without rotation of circular cylinder and was studied by using boundary layer techniques which has two major deficiencies: one is that it is valid for high Reynolds number flow and not for low *R* and secondly solution or integration of constitutive equations no longer converges for large values of time t. The later type of difficulty has been encountered by various authors, for example [6] and [37-41] for Newtonian fluids. To overcome these difficulties we adopt higher order numerical scheme which is valid for all Reynolds number rotation parameter, and especially for all time. The detail of this scheme is given in the following section. This numerical scheme reduces the highly non-linear system of partial differential equations to system of difference equations, which then will be solved by the SOR-iterative method, which has capability to accelerate the convergence of the iterative scheme. Henceforth this numerical procedure is efficient, for studies of such type of sensitive flow problems and is straightforward/ economical in core storage requirements of a computer and easy to programme.

## *4. Numerical Scheme*

We shall use the following notation. The grid size along s-, θ-, and t-directions, will be taken by H, $K_1$, $K_2$ respectively, and the points $(s_0, \theta_0, t_0)$, $(s_0+H, \theta_0, t_0)$, $(s_0, \theta_0+K_1, t_0)$, $(s_0-H, \theta_0, t_0)$, $(s_0, \theta_0-K_1, t_0)$, $(s_0, \theta_0, t_0+K_2)$, $(s_0, \theta_0, t_0-K_2)$, $(s_0+H, \theta_0, t_0-K_2)$, and $(s_0-H, \theta_0, t_0-K_2)$ will be represented by the subscripts 0, 1, 2, 3, 4, 5, 6, 16, and 36 respectively. Moreover, the grid which we adopted to discreatize the whole domain of computation is given in figure 1a.
This scheme consists of two methods, which are described as follows:

### *Adams-Bashforth Temporal Fourier Series Method (ABTFSM)*

In this method we consider

$$\psi = \sum_{n=1}^{\infty} f_n(s,t) \sin n\theta \qquad (4.1)$$

$$E = \sum_{n=1}^{\infty} g_n(s,t) \sin n\theta \qquad (4.2)$$

$$\eta = \sum_{n=1}^{\infty} h_n(s,t) \sin n\theta \qquad (4.3)$$

On using equations (4.1) - (4.3) into equation (3.1) – (3.3), multiplying the equation by *sinnθ*, and integrating with respect to θ from *θ*=0 to *θ*=π, we obtain

$$\frac{\partial^2 f_n}{\partial s^2} - n^2 f_n = -e^{2s} g_n, \qquad (4.4)$$

$$e^{2s}\frac{\partial g_n}{\partial t} = \frac{2}{R}\frac{\partial^2 g_n}{\partial s^2} + nf_{2n}\frac{\partial g_n}{\partial s} + \left[\frac{n}{2}\frac{\partial f_{2n}}{\partial s} - \frac{2n^2}{R}\right]g_n - \frac{C_1}{R}\frac{\partial^2 h_n}{\partial s^2} + \frac{n^2 C_1 h_n}{R} + P_n \quad (4.5)$$

$$e^{2s}\frac{\partial h_n}{\partial t} = \frac{1}{C_3}\frac{\partial^2 h_n}{\partial s^2} + nf_{2n}\frac{\partial h_n}{\partial s} + \left[\frac{n}{2}\frac{\partial f_{2n}}{\partial s} - \frac{2C_2 e^{2s}}{C_3} - \frac{n^2}{C_3}\right]g_n + \frac{2C_2 e^{2s} g_n}{C_3} + Q_n \quad (4.6)$$

where

$$P_n = \frac{1}{2}\sum_{\substack{m=1 \\ (m \neq n)}}^{\infty}\left[\{(m+n)f_{m+n} - qf_q\}\frac{\partial g_m}{\partial s} + m\left\{\frac{\partial f_{m+n}}{\partial s} - sgn(m-n)\frac{\partial f_q}{\partial s}\right\}g_m\right] \quad (4.7)$$

and

$$Q_n = \frac{1}{2}\sum_{\substack{m=1 \\ (m \neq n)}}^{\infty}\left[\{(m+n)f_{m+n} - qf_q\}\frac{\partial h_m}{\partial s} + m\left\{\frac{\partial f_{m+n}}{\partial s} - sgn(m-n)\frac{\partial f_q}{\partial s}\right\}h_m\right] \quad (4.8)$$

and $q = |m-n|$ and *sgn(m-n)* represents the sign of m-n with *sign(0)=0*. In equations (4.1) – (4.8), n is taken as positive integer and the solutions of these equations define, in theory, three infinite sets of equations in $f_n(s, t)$ $g_n(s, t)$, and $h_n(s, t)$. In practice, the series in equations (4.1) – (4.3) are truncated to $N_0$. All functions with subscript $n > N_0$ are taken zero in the expansion of Eqs. (4.1)–(4.8). In terms of $f_n$, $g_n$, and $h_n$, the boundary conditions given in (3.4) would deform as: and the boundary conditions given in equation (3.5) in terms of $f_n$ and $g_n$ become

$$f_n = 0, \quad \frac{\partial f_n}{\partial s} = \frac{2\alpha(\cos n\pi - 1)}{n\pi}, \quad h_n = 0 \quad \text{on } s = 0, \forall n, \text{ and } \forall t \quad (4.9)$$

where $\delta_1 = 1$ while $\delta_n = 0$ for n=2, 3, 4, …………..

$$e^{-s}\frac{\partial f_n}{\partial s} \to t\delta_n \quad \text{and} \quad e^{-s}f_n \to t\delta_n \qquad \text{as} \quad s \to \infty \quad (4.10)$$

$$\text{and} \qquad g_n \to 0, h_n \to 0 \qquad \text{as} \quad s \to \infty \quad (4.11)$$

$$\int_0^{\infty} e^{(2-n)s}g_n(s,t)ds = 2\delta_n \quad (4.12)$$

An alternative form of boundary condition given in (4.10) can be obtained by multiplying equation (3.1) by $e^{-ns}$ and integrating from *s*=0 to *s*=∞ and then using (4.9) and (4.11) we get
Another explicit form of (4.10) can be obtained by using (4.4), (4.9), and (4.11) which is

$$f_n \to 2t\sinh s\, \delta_n \qquad \text{as } s \to \infty \quad (4.13)$$

Finally the boundary condition for $g_n$ over the surface of a cylinder will take the form after using (4.1) – (4.3), as

$$(g_n)_w = -\frac{[6(f_n)_1 + H^2(g_n)_1]}{2[H^2 + H^3]} + \frac{3\alpha[\cos(n\pi) - 1]}{nH[H+1]} \quad (4.14)$$

Now the boundary conditions given above are sufficient to solve equations (4.4) to (4.8). Next to transform the set of partial differential equations given in equations (4.4)–(4.8) into difference equations, we use the second-order Adams-Bashforth temporal method and central difference for spatial coordinates.
Consequently the equations to be solved after approximating at point "0", are

$$-n^2 H^2 (f_n)_0 + (f_n)_1 + (f_n)_3 + H^2 e^{2s_0}(g_n)_0 = 0 \quad (4.15)$$

$$(A_n)_0(g_n)_0 + (B_n)_0(g_n)_3 - (g_n)_5 + (D_n)_6(g_n)_6 - (L_n)_0(g_n)_0 + (N_n)_0(h_n)_0$$
$$+ (M_n)_0[(h_n)_1 + (h_n)_3] + (R_n)_0(P_n)_0 + (T_n)_6(g_n)_{16} + (U_n)_6(g_n)_{36}$$
$$- \frac{1}{3}(M_n)_0[(h_n)_{16} + (h_n)_{36}] - \frac{1}{3}(N_n)_0(h_n)_6 - \frac{1}{3}(R_n)_0(P_n)_6 = 0 \qquad (4.16)$$

and

$$(A'_n)_0(h_n)_0 + (B'_n)_0(h_n)_1 + (D'_n)_0(h_n)_3 - (h_n)_5 + (I_n)_6(h_n)_6$$
$$+ (J_n)_6(h_n)_{16} + (K_n)_6(h_n)_{36} + (O_n)_0(g_n)_0 + (R_n)_0(Q_n)_0$$
$$- \frac{1}{3}(O_n)_0(g_n)_6 - \frac{1}{3}(R_n)_0(Q_n)_6 = 0 \qquad (4.17)$$

*where*

$$(A_n)_0 = \frac{3e^{-2s_0}K_2}{2}\left[\frac{2}{RH^2} + \frac{n}{2H}(f_{2n})_0\right]$$

$$(B_n)_0 = \frac{3e^{-2s_0}K_2}{2}\left[\frac{2}{RH^2} - \frac{n}{2H}(f_{2n})_0\right]$$

$$(A'_n)_6 = 1 + \frac{3e^{-2s_0}K_2}{2}\left[\frac{-2}{C_3H^2} - \frac{n^2}{C_3} - \frac{2C_2}{C_3}e^{2s_0} + \frac{n}{4H}\{(f_{2n})_1 - (f_{2n})_3\}\right]$$

$$(B'_n)_6 = \frac{3e^{-2s_0}K_2}{2}\left[\frac{1}{C_3H^2} + \frac{n}{2H}(f_{2n})_0\right]$$

$$(D_n)_6 = \frac{e^{-2s_0}K_2}{2}\left[\frac{-4}{RH^2} - \frac{2n^2}{R} + \frac{n}{4H}\{(f_{2n})_{16} - (f_{2n})_{36}\}\right]$$

$$(D'_n)_6 = \frac{3e^{-2s_0}K_2}{2}\left[\frac{1}{C_3H^2} - \frac{n}{2H}(f_{2n})_0\right]$$

$$(I_n)_6 = -\frac{e^{-2s_0}K_2}{2}\left[\frac{-2}{C_3H^2} - \frac{n^2}{C_3} - \frac{2C_2}{C_3}e^{2s_0} + \frac{n}{4H}\{(f_{2n})_{16} - (f_{2n})_{36}\}\right]$$

$$(J_n)_6 = \frac{-e^{-2s_0}K_2}{2}\left[\frac{1}{C_3H^2} + \frac{n}{2H}(f_{2n})_6\right]$$

$$(K_n)_6 = \frac{-e^{-2s_0}K_2}{2}\left[\frac{1}{C_3H^2} - \frac{n}{2H}(f_{2n})_6\right]$$

$$(L_n)_0 = 1 + \frac{3e^{-2s_0}K_2}{2}\left[\frac{-4}{RH^2} - \frac{2n^2}{R} + \frac{n}{4H}\{(f_{2n})_1 - (f_{2n})_3\}\right]$$

$$(M_n)_0 = \frac{3e^{-2s_0}K_2C_1}{2RH^2}, \qquad (N_n)_0 = \frac{3e^{-2s_0}K_2C_1}{2RH^2}[2 + n^2H^2]$$

$$(O_n)_0 = \frac{3C_2 K_2}{C_3}, \qquad (R_n)_0 = \frac{3e^{-2s_0} K_2}{2}$$

$$(T_n)_6 = \frac{-e^{-2s_0} K_2}{2}\left[\frac{2}{RH^2} + \frac{n}{2H}(f_{2n})_6\right]$$

$$(U_n)_6 = \frac{-e^{-2s_0} K_2}{2}\left[\frac{2}{RH^2} - \frac{n}{2H}(f_{2n})_6\right]$$

### *Runge-Kutta Temporal Special Finite Difference Method*

Let us split equation (3.2) into following three equations

$$\frac{\partial^2 E}{\partial s^2} + B\frac{\partial E}{\partial s} = A \tag{4.18}$$

$$\frac{\partial^2 E}{\partial \theta^2} + C\frac{\partial E}{\partial \theta} = -\frac{A}{2} \tag{4.19}$$

and $\quad C_1\left[\dfrac{\partial^2 \eta}{\partial s^2} + \dfrac{\partial^2 \eta}{\partial \theta^2}\right] + \text{Re}^{2s}\dfrac{\partial E}{\partial t} = A \tag{4.20}$

Now, on following the procedure of the method given in [31], we can write:

$$[1 + \frac{B_0 H}{2} + \frac{B_0^2 H^2}{8}]E_1 + [\frac{H^2}{K_1^2}\{1+\frac{K_1^2 C_0^2}{8}\} + \frac{C_0 H^2}{2K_1}]E_2 + [1 - \frac{B_0 H}{2} + \frac{B_0^2 H^2}{8}]E_3$$

$$+ [\frac{H^2}{K_1^2}\{1+\frac{K_1^2 C_0^2}{8}\} - \frac{C_0 H^2}{2K_1}]E_4 - \frac{C_1}{2}[\eta_1 + \eta_3] - [2 + \frac{2H^2}{K_1^2} + \frac{B_0^2 H^2}{4} + \frac{C_0^2 H^2}{4}]E_0$$

$$- \frac{C_1 H^2}{2K_1^2}[\eta_2 + \eta_4] + \frac{C_1}{H^2}[H^2 + K_1^2]\eta_0 + L_0 H^2\left(\frac{\partial E}{\partial t}\right)_0 = 0 \tag{4.21}$$

Alternatively, the above equation can be written as:

$$\frac{L_0 H^2}{a_0}\left(\frac{\partial E}{\partial t}\right)_0 = \tilde{Z}(E, \eta, t) \tag{4.22}$$

where $\quad \tilde{Z}(E, t) = \sum_{i=0}^{4} C'_i E_i + C'_5[\eta_1 + \eta_3] + C'_6[\eta_2 + \eta_4] + C'_7 \eta_0 \tag{4.23}$

and

$$C'_1 = -\frac{[1 + \dfrac{B_0 H}{2} + \dfrac{B_0^2 H^2}{8}]}{a_0}, \qquad C'_2 = -\frac{[\dfrac{H^2}{K_1^2}\{1+\dfrac{K_1^2 C_0^2}{8}\} + \dfrac{C_0 H^2}{2K_1}]}{a_0},$$

$$C'_3 = -\frac{[1 - \dfrac{B_0 H}{2} + \dfrac{B_0^2 H^2}{8}]}{a_0}, \qquad C'_4 = -\frac{[\dfrac{H^2}{K_1^2}\{1+\dfrac{K_1^2 C_0^2}{8}\} - \dfrac{C_0 H^2}{2K_1}]}{a_0},$$

$$C'_5 = \frac{C'_1}{a_0}, \qquad\qquad C'_6 = \frac{H^2}{K_1^2} C'_5,$$

$$C'_7 = -2(C'_5 + C'_6), \qquad\qquad L = -\frac{R}{2} e^{2s},$$

$$a_0 = 2 + 2\frac{H^2}{K_1^2} + \frac{H^2}{4}(B_0^2 + C_0^2), \qquad C_0 = 1.$$

Next, on following a similar procedure for equation (3.3) used for equation (2.12), we get

$$\frac{\tilde{L}_0 H^2}{b_0}\left(\frac{\partial E}{\partial t}\right)_0 = \tilde{\tilde{Z}}(E, \eta, t) \qquad (4.24)$$

where $\quad \tilde{\tilde{Z}}(E, t) = \sum_{i=0}^{4} D_i \eta_i + D_5 E_0 \qquad (4.25)$

$$D_1 = -\frac{[1 + \frac{U_0 H}{2} + \frac{U_0^2 H^2}{8}]}{a_0}, \qquad D_2 = -\frac{[\frac{H^2}{K_1^2}\{1 + \frac{K_1^2 V_0^2}{8}\} + \frac{V_0 H^2}{2K_1}]}{a_0}$$

$$D_3 = -\frac{[1 - \frac{U_0 H}{2} + \frac{U_0^2 H^2}{8}]}{a_0}, \qquad D_4 = -\frac{[\frac{H^2}{K_1^2}\{1 + \frac{K_1^2 V_0^2}{8}\} - \frac{V_0 H^2}{2K_1}]}{a_0}$$

$$D_5 = -2H^2 C_2 e^{2s}, \qquad\qquad \tilde{L} = \frac{2C_3 L}{R}$$

$$U = \frac{2BC_3}{R}, \qquad\qquad V = \frac{2CC_3}{R}$$

$$b_0 = 2 + 2\frac{H^2}{K_1^2} + \frac{H^2}{4}(U_0^2 + V_0^2) + 2H^2 C_2 e^{2s_0}, \qquad D_0 = 1$$

$$B = -\frac{R}{2}\frac{\partial \psi}{\partial \theta}, \qquad C = \frac{R}{2}\frac{\partial \psi}{\partial s},$$

$$U = -C_3 \frac{\partial \psi}{\partial \theta}, \text{ and } V = C_3 \frac{\partial \psi}{\partial s}$$

Then using 'Runge-Kutta temporal special finite-difference method' for equations (4.22) and (4.24) we found that,

$$E_5 = E_0 + \frac{K_2 a_0}{6 L_0 H^2}\left[\tilde{a} + 2\tilde{b} + 2\tilde{d} + \tilde{e}\right] \qquad (4.26)$$

$$\eta_5 = \eta_0 + \frac{K_2 b_0}{6 \tilde{L}_0 H^2}\left[\tilde{\tilde{a}} + 2\tilde{\tilde{b}} + 2\tilde{\tilde{d}} + \tilde{\tilde{e}}\right] \qquad (4.27)$$

where $\tilde{a} = \tilde{Z}(E_0, \eta_0, t_0)$

$\tilde{b} = \tilde{Z}(E_0 + \frac{K_2\tilde{a}}{2}, \eta_0 + \frac{K_2\tilde{a}}{2}, t_0 + \frac{K_2}{2})$

$\tilde{d} = \tilde{Z}(E_0 + \frac{K_2\tilde{b}}{2}, \eta_0 + \frac{K_2\tilde{b}}{2}, t_0 + \frac{K_2}{2})$

$\tilde{e} = \tilde{Z}(E_0 + K_2\tilde{d}, \eta_0 + K_2\tilde{d}, t_0 + K_2)$

$\tilde{\tilde{a}} = \tilde{\tilde{Z}}(E_0, \eta_0, t_0)$

$\tilde{\tilde{b}} = \tilde{\tilde{Z}}(E_0 + \frac{K_2\tilde{\tilde{a}}}{2}, \eta_0 + \frac{K_2\tilde{\tilde{a}}}{2}, t_0 + \frac{K_2}{2})$

$\tilde{\tilde{d}} = \tilde{\tilde{Z}}(E_0 + \frac{K_2\tilde{\tilde{b}}}{2}, \eta_0 + \frac{K_2\tilde{\tilde{b}}}{2}, t_0 + \frac{K_2}{2})$

and $\tilde{\tilde{e}} = \tilde{\tilde{Z}}(E_0 + K_2\tilde{\tilde{d}}, \eta_0 + K_2\tilde{\tilde{d}}, t_0 + K_2)$

Moreover, the value of $\psi$ in equations (4.22) and (4.24) can be obtained on solving the equation (3.1).

## 5. *Boundary Condition on "E" at large "s"------------A higher order formula*

In this section, we shall present another technique for the solution of governing equations (3.1) – (3.3) at large s, which is briefly given as follow. Let us represent the point ((N+1)H, θ₀, t₀), (NH, θ₀, t₀), ((N-1)H, θ₀, t₀), ((N-2)H, θ₀, t₀), ((N-3)H, θ₀, t₀), ((N+1)H, θ₀+K₁, t₀), ((N+1)H, θ₀-K₁, t₀), ((N+1)H, θ₀, t₀+K₂), and ((N+1)H, θ₀, t₀-K₂) by subscripts a, b, c, d, e, f, g, h, and i respectively. On using the boundary condition on $\psi$ at infinity into equation (3.2), we get

$$\left(\frac{\partial^2 E}{\partial s^2}\right)_a + \left(\frac{\partial^2 E}{\partial \theta^2}\right)_a + \left(\frac{Rte^s \sin(\theta)}{2} \frac{\partial E}{\partial \theta}\right)_a - \left(\frac{Rte^s \cos(\theta)}{2} \frac{\partial E}{\partial s}\right)_a - \frac{C_1}{2}\left(\frac{\partial^2 \eta}{\partial s^2} + \frac{\partial^2 \eta}{\partial \theta^2}\right)_a$$
$$= \left(C_3 e^{2s} \frac{\partial \eta}{\partial t}\right)_a \quad (5.1)$$

The above equation will be approximated at the point "a" by the method similar to that of special finite-difference method we obtain

$$\left(\frac{\partial^2 f}{\partial s^2}\right)_a = \frac{1}{12H^2}\left[-104 f_b + 114 f_c - 56 f_d + 11 f_e + 35 f_a\right] + O(H^3) \quad (5.2)$$

and after some simplifications we get the following implicit scheme for the boundary condition at infinity that

$$\left[35 - \frac{24H^2}{K_1^2} - 3H^2\left(P_a^2 + Q_a^2\right)\right]E_a - 104\left[1 - \frac{HP_a}{2} + \frac{H^2U_a}{8} - \frac{H^3V_a}{96}\right]E_b$$

$$+ 104\left[1 - HP_a + \frac{H^2U_a}{8} - \frac{H^3S_a}{96}\right]E_c - 56\left[1 - \frac{3HP_a}{2} + \frac{9H^2U_a}{8} - \frac{9H^3V_a}{32}\right]E_d$$

$$+ 11\left[1 - 2HP_a + 2H^2U_a - \frac{4H^3S_a}{3}\right]E_e + \frac{12H^2}{K_1^2}\left[1 + \frac{K_1Q_a}{2} + \frac{K_1^2}{8}\left(Q_a^2 - 2P_a\right)\right]E_f$$

$$+ \frac{12H^2}{K_1^2}\left[1 - \frac{K_1Q_a}{2} + \frac{K_1^2}{8}\left(Q_a^2 - 2P_a\right)\right]E_g - \frac{35}{2}C_1\eta_a + 52C_1\eta_b - 57C_1\eta_c + 28C_1\eta_d$$

$$- \frac{11}{2}C_1\eta_e - \frac{6H^2C_1}{K_1^2}\left[\eta_f + \eta_g\right] = \frac{6H^2L_a}{K_2}\left[E_i - E_h\right] \quad (5.3)$$

where

$$P = -\frac{R}{2}te^s \cos(\theta), \qquad Q = \frac{R}{2}te^s \sin(\theta), \qquad U = P^2 + 2P,$$

$$V = P^3 + 12P^2 + 8P, \qquad S = P^3 + 6P^2 + 4P, \quad \text{and} \quad L = -\frac{R}{2}e^{2s}$$

The variation of this non-zero spin has been simulated for $\alpha=0$ and is given in the section 8 and graphically in figure 21.

## 6. *Boundary Condition on "$\eta$" at large "s"------------A higher order formula*

Analogous to section 5, let us use infinite boundary condition on $\psi$ into equation (3.3), we get,

$$\left(\frac{\partial^2\eta}{\partial s^2}\right)_a + \left(\frac{\partial^2\eta}{\partial \theta^2}\right)_a + \left(\frac{Rte^s \sin(\theta)}{2}\frac{\partial\eta}{\partial\theta}\right)_a - \left(\frac{Rte^s \cos(\theta)}{2}\frac{\partial\eta}{\partial s}\right)_a - 2C_2 e^{2s}(E - \eta)_a$$

$$= \left(C_3 e^{2s}\frac{\partial\eta}{\partial t}\right)_a \quad (6.1)$$

Following the similar procedure for equation (6.1) of special finite-difference method, we get

$$\left[35 - \frac{24H^2}{K_1^2} - 3H^2\left(\tilde{P}_a^2 + \tilde{Q}_a^2\right)\right]\eta_a - 104\left[1 - \frac{H\tilde{P}_a}{2} + \frac{H^2\tilde{U}_a}{8} - \frac{H^3\tilde{V}_a}{96}\right]\eta_b$$

$$+ 104\left[1 - H\tilde{P}_a + \frac{H^2\tilde{U}_a}{8} - \frac{H^3\tilde{S}_a}{96}\right]\eta_c - 56\left[1 - \frac{3H\tilde{P}_a}{2} + \frac{9H^2\tilde{U}_a}{8} - \frac{9H^3\tilde{V}_a}{32}\right]\eta_d$$

$$+ 11\left[1 - 2H\tilde{P}_a + 2H^2\tilde{U}_a - \frac{4H^3\tilde{S}_a}{3}\right]\eta_e + \frac{12H^2}{K_1^2}\left[1 + \frac{K_1\tilde{Q}_a}{2} + \frac{K_1^2}{8}\left(\tilde{Q}_a^2 - 2\tilde{P}_a\right)\right]\eta_f$$

$$+ \frac{12H^2}{K_1^2}\left[1 - \frac{K_1\tilde{Q}_a}{2} + \frac{K_1^2}{8}\left(\tilde{Q}_a^2 - 2\tilde{P}_a\right)\right]E_g + 24C_2H^2 e^{2s_0}E_a = \frac{6H^2L_a}{K_2}\left[E_i - E_h\right] \quad (6.2)$$

where

$$\tilde{P} = -C_3 t e^s \cos(\theta), \qquad \tilde{Q} = C_3 t e^s \sin(\theta), \qquad \tilde{U} = P^2 + 2P,$$
$$\tilde{V} = \tilde{P}^3 + 12\tilde{P}^2 + 8\tilde{P}, \qquad \tilde{S} = \tilde{P}^3 + 6\tilde{P}^2 + 4\tilde{P}, \quad \text{and} \quad \tilde{L} = -C_3 e^{2s}$$

The variation of this non-zero spin has been simulated for α=0 and is given in the section 8 and graphically in figure 21.

## 7. *Computational Procedure*

We present the brief survey of computational procedure here:

- Initially equations (4.15) – (4.17) are solved with respect to the boundary conditions given in (4.9), (4.10) (or (4.12) or (4.13) ), and (4.14) till some time $t_c$ (where $t_c$ is the final time limit at which the method to converge corresponding to R). From this we get the solutions $f_n$, $g_n$, and $h_n$. These $f_n$, $g_n$, and $h_n$ are substituted in equaitons (4.1) – (4.3) to get ψ, E, and η.
- In the second step, at time $t=t_c$, the values of ψ, E, and η (of $t \leq t_c$) are further used as initial estimates of higher order RKTSFD method to solve equations (4.26) and (4.27) with respect to boundary conditions (3.4) and (3.5).

We observe that this numerical scheme increases the rate of convergence and reduces the time of execution for micropolar fluid also. So this scheme is economical and fast. In the above procedure we use the point S. O. R- iterative procedure [42], subject to the appropriate boundary conditions.

Initially the programme is executed with boundary conditions "E=η=0 as s→∞", when some accuracy is achieved, then these boundary conditions are disconnected and the boundary conditions given in equations (5.3), and (6.2) are used. The above procedure is repeated until convergence is obtained according to the criterion,

$$\max \left| E_0^{(m+1)} - E_0^{(m)} \right| < 10^{-5}, \; \max \left| \eta_0^{(m+1)} - \eta_0^{(m)} \right| < 10^{-5} \; \text{and} \; \max \left| \psi_0^{(m+1)} - \psi_0^{(m)} \right| < 10^{-5}$$

where the superscript 'm' represents the number of iteration. The results obtained are further refined and enhanced up to the sixth order by Richardson's extrapolation method [43]. Consequently, this scheme is valid for all *R* and *t* with high rate of convergence.

## 8. *Calculated Results and Discussion*

The solutions have been obtained for R∈{1, 10, 100, 200, 349.285} with grid sizes H ∈{1/20, 1/40, 1/60, 1/80, 1/160} and $K_1$∈{π/40, π/80, π/160} such that $C_i$,'s are given in table 1.

| *Cases* | $C_1$ | $C_2$ | $C_3$ |
|---|---|---|---|
| **Case 1** | 0.1 | 0.75 | 0.1 |
| **Case 2** | 1.5 | 0.9 | 2.3 |
| **Case 3** | 2.5 | 3.0 | 0.5 |

**Table 1: Three cases for adopting the values of constants**.

For low and moderate values of *R* the value of $K_2$ is taken as 1/130 while for high value of *R* it is taken as 1/230, and this numerical experiment is done up to *t*=80 but very special attention is focused on flow in the interval 0<*t*≤1 for flow development, generation and variation of secondary vortices in the vicinity of the circular cylinder. As the flow begins to start, the cylinder has to experience some extra forces which may reduce as the time passes but its behaviour is different as rotation parameter α is increased. That is why, we take $K_2=10^{-4}$ when the integration is started in all cases and for low values of time. This numerical experiment is performed for various values of α in the range 0≤α≤4.5, attention is focused on four values of α namely α∈{0, 2.5, 3, 4.5} for simulation and discussion purposes.

Let us introduce four parameters $\psi_{max}$, $\psi_{min}$, $E_{min}$, and $\eta_{min}$ for comparisons of two aforementioned methods mutually and with those of Newtonian one by taking χ=η=0 in the governing equations for micropolar fluids. Here $\psi_{max}$, $\psi_{min}$, $E_{min}$, and $\eta_{min}$ represent for maximum value of stream function, maximum value of vorticity, minimum value of vorticity and minimum value of spin respectively. The comparison of results is presented in tables 2 and 3.

We split the discussion in two cases regarding the symmetric and asymmetric flow configuration since rotation of the circular cylinder causes the asymmetric behaviour of the flow.

**Case (a). Symmetric Flow**

In this case, the flow is supposed to be symmetric for all $R$, $\alpha$ and $t$ and therefore we consider upper half of the cylinder as the domain of computation.

One of the interesting features of the present problem is the question of the growth of the standing vortices behind the cylinder as Reynolds number $R$, rotation parameter α and time $t$ vary. In figures 2–9, streamlines are displayed for the flow past the upper half of the cylinder for different values of R and α as mentioned above at four selected time levels namely t∈{0.046, 0.1, 0.154, 0.993}. Figure 2 shows the temporal development of the flow for $R=1$, $\alpha=2.5$ by results obtained on solving the equations by ABTFSM. The formation of secondary flow with closed streamlines first appear at $t=0.993$ on the surface of the cylinder even when $R$ is very low, i.e. $R=1$, and when the cylinder is not rotating, as shown in figure 3(d). This fact is only related to micropolar fluid for this problem since this is not so for Newtonian one. When $R$ is moderate and α=0 then secondary eddy will appear in the wake of the cylinder. This eddy will also appear as earlier as $R$ converges to higher values and it grows steadily as $R$ increases. If $d$ represents the length of this vortex pair then we can observe that it increases as $R$ increases at any instant of time. Note that this was the case when there is no rotation of the cylinder. If the rotation of the cylinder is considered and the cylinder starts to rotate also in addition to the accelerated translation, then the flow-behaviour is slightly different. The appearance of the eddy in the wake is delayed to successively higher times as α increases and it appears after $t>1$. But some portion of the fluid which is adhere on the surface of the cylinder will starts to rotate along with the rotation of the cylinder and this section of the fluid can be seen as closed streamlines in the vicinity of the cylinder in figures 4–9. These closed streamlines has characteristics that they dissolve the vortex pair and hence Karman vortex street appears around the cylinder in early stage of the flow for all values of $R$ and $\alpha$. These closed streamlines around the cylinder are rotating counterclockwise around the cylinder and its thickness decreases as R increases as well as time passes. The main effects of the rotation of the cylinder is that it causes the delay in the appearance of eddies but also it controls the boundary layer separation as shown in figures 4–9. For higher values of both $R$ and $\alpha$ the situation is different. The high speed of accelerated flow dominates and reduces the rotation effects and the vortices in the wake and will appear at the front stagnation point within the closed streamlined circulating secondary flow region in very early stage of the flow development as shown in figure 7(a). These eddies will overlap and form a large diameter eddy and shift to front stagnation point as rotation speed is further increased as shown in figure 9.

The growth of boundary layer thickness and boundary layer separation are also examined by studying the equi-vorticity lines, the variation of the vorticity over the surface of the cylinder with respect to θ the variation of $E_{max}$ and $E_{min}$ for various values of $R$, $\alpha$, and $t$.
Here as a sample, equi-vorticity lines for $\alpha \in \{0, 2.5, 4.5\}$ when $R=10$ at four time levels are displayed in figures 10 and 11. These figures show that the boundary layer thickness decreases as $R$ increases and increases as time passes when either the cylinder is rotating or not. Moreover, it also decreases if rotation is increased. Figures 17 and 18 indicate that (i) the separated region increases as time increases before the rotation but it decreases as rotation increases; (ii) the separation point occurs sometimes after $t=1$ when α is zero while this time-limit will delay separation when $\alpha$ is not equal to zero.

Equi-spin lines characterize the motion of the individual particles of the fluid. Equi-spin lines are drawn for all cases as mentioned above but here some figures 12 and 13 are given for R∈{1, 10, 100} and α∈{0, 2.5, 3, 4.5}. These figures show that the rotation reduces the micropolar spin boundary layer. Figure 12 represents the variation of equi-spin lines for $R=100$ when the cylinder is not rotating at time instants mentioned above. The numerous peaks in this figure correspond to the multiplication of spin vortices and there exist more than one separation point. But when rotation is imposed, these multi-separation points coincide to a single point and this fluctuations in spin is also studied by plotting $\eta_{min}$ versus time in the range 0≤t≤1 which is presented in figure 22. This figure indicates that when $\alpha=0$, the numerous peaks of ($\eta_{min}$, t) curve correspond to the sudden change in spin happened in early stage of the flow but it is not so and it becomes smooth and calm variation when $\alpha$ is not equal to zero. This variation remains same for all non-zero values of $\alpha$ for a fixed value of $R$. The variation in surface forces, are also examined for all cases as mentioned above but here figures 14 and 15 are presented for the temporal development in drag and lift over the surface of the cylinder for $\alpha \in \{0, 2.5, 3, 4.5\}$ when $R=100$ respectively. On studying these it will be seen that for low values of $\alpha$ the drag remains constant and very low but for moderate and slightly high value of $\alpha$, it fluctuates and oscillate even in very early stage of the start of the flow for all R as shown in figure 14. Figure 15 shows the temporal variation of lift coefficient, which is vice versa of the variation of drag coefficient.

The temporal variation in the pressure ($P_1$) over the surface of the cylinder before and after the rotation of the cylinder is also studied and is presented in figure 16. This pressure is estimated by the formula given as follows:

$$P_1(\theta, t) = t^2 - \alpha^2 - \frac{4}{R}\int_{\theta_0}^{\theta}\left(\frac{\partial E}{\partial s} + C_1\frac{\partial \xi}{\partial s}\right)_{s=0} d\theta - \frac{4}{R}\int_{0}^{\infty}\left(\frac{\partial E}{\partial s} + C_1\frac{\partial \xi}{\partial s}\right)_{\theta=\theta_0} ds$$

$$+ \int_{0}^{\infty} 2\left(\frac{\partial^2 \psi}{\partial \theta \partial t} - E\frac{\partial \psi}{\partial s}\right)_{\theta=\theta_0} ds$$

*For the range $0 \leq \alpha \leq 2.5$,*
- the pressure over the surface of the surface of cylinder decreases as time passes.
- this pressure increases steadily in the direction from rear stagnation point to the front stagnation point of the cylinder.
- the minimum pressure over the surface of the cylinder exists at rear stagnation point of the cylinder.
- the maximum pressure over the surface occurs near the front stagnation point of the cylinder.

as shown in figure 16(a-b).

*For $\alpha > 2.5$,* the pressure over the surface of the cylinder fluctuates as time passes. Its extreme values vary with time as shown in figures 16(c-d).

**Case (b). Asymmetric Flow**

In this case, the flow is supposed to be asymmetric for all $R$, $\alpha$ and $t$ and therefore we consider complete domain of computation with $0 \leq \theta \leq 2\pi$, which is shown in figure 1b.

The calculations are carried out for all values of flow, fluid, and geometric parameters, which are mentioned in case (a). Here all the simulations will not be displayed but as a sample we present figures 19 and 20. For low values of Reynolds number e.g. $R=1$, and low values of $\alpha$ with $\alpha < 4.5$, it is seen that flow is symmetric. For $R=1$, $\alpha \geq 4.5$ the flow becomes asymmetric. The temporal development of the flow behaviour for $R=1$, $\alpha=4.5$ is shown in figure 19. Moreover, the temporal growth of equi-vorticity lines for this case is also simulated in figure 20.

## 9. Conclusions

In this study, the enhancement of lift and reduction in drag was observed if the micropolarity effects are intensified. Same is happened if the rotation of a cylinder increases. Furthermore, the vortex-pair in the wake is delayed to successively higher times as rotation parameter $\alpha$ increases. In addition, the rotation helps not only in dissolving vortices adjacent to the cylinder and adverse pressure region but also in dissolving the boundary layer separation. Furthermore, the rotation reduces the micropolar spin boundary layer also.

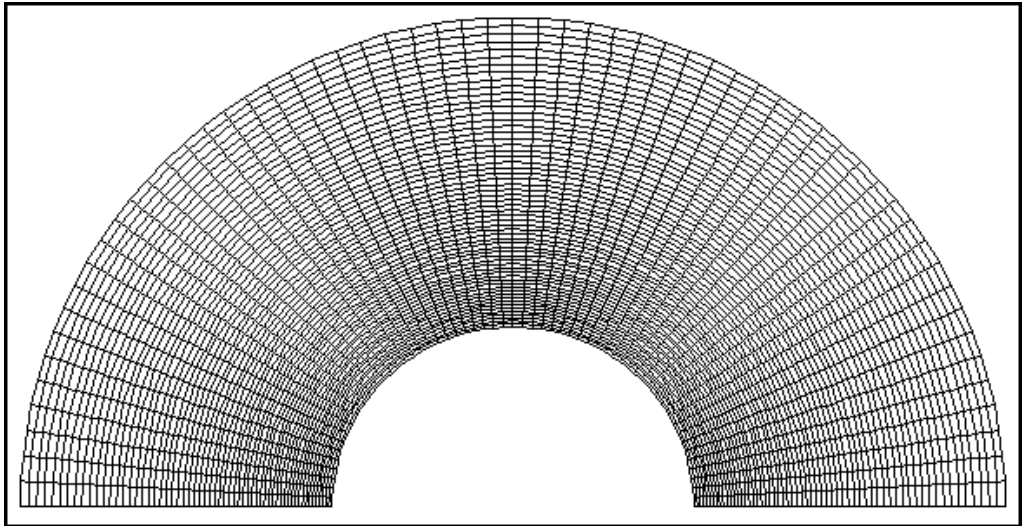
(a)

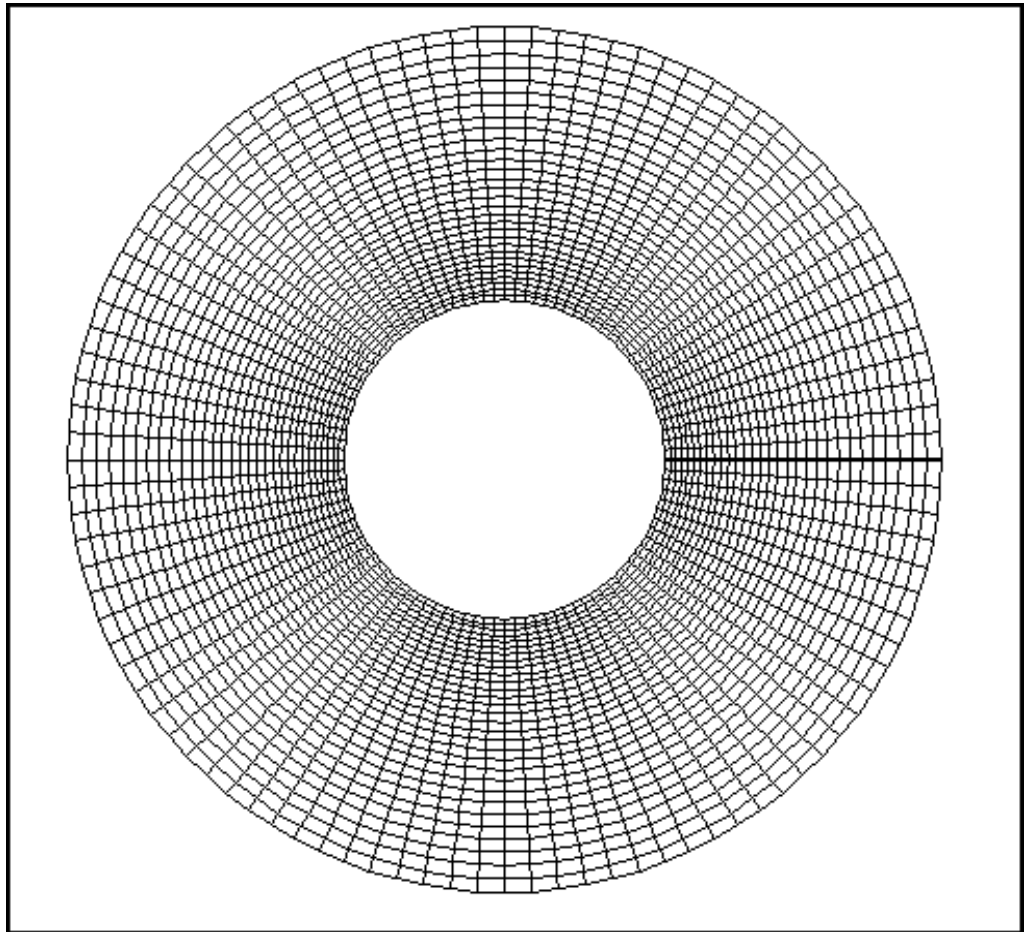
(b)

**Fig.1. The meshing of the computational domain for (a) symmetric flow (b)asymmetric flow.**

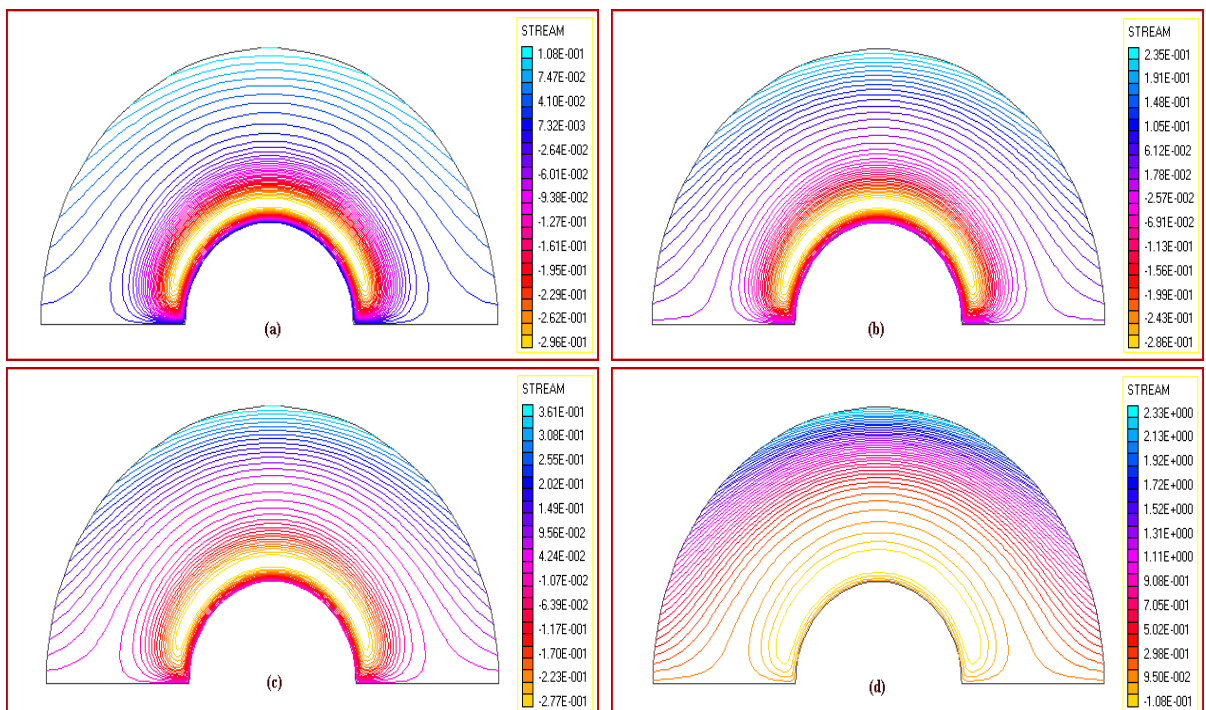

**Fig. 2. Streamlines for R=1, α=2.5 obtained by ABTFS method at time levels (a) t=0.046 (b) t=0.1 (c) t=0.154 (d) t=0.993.**

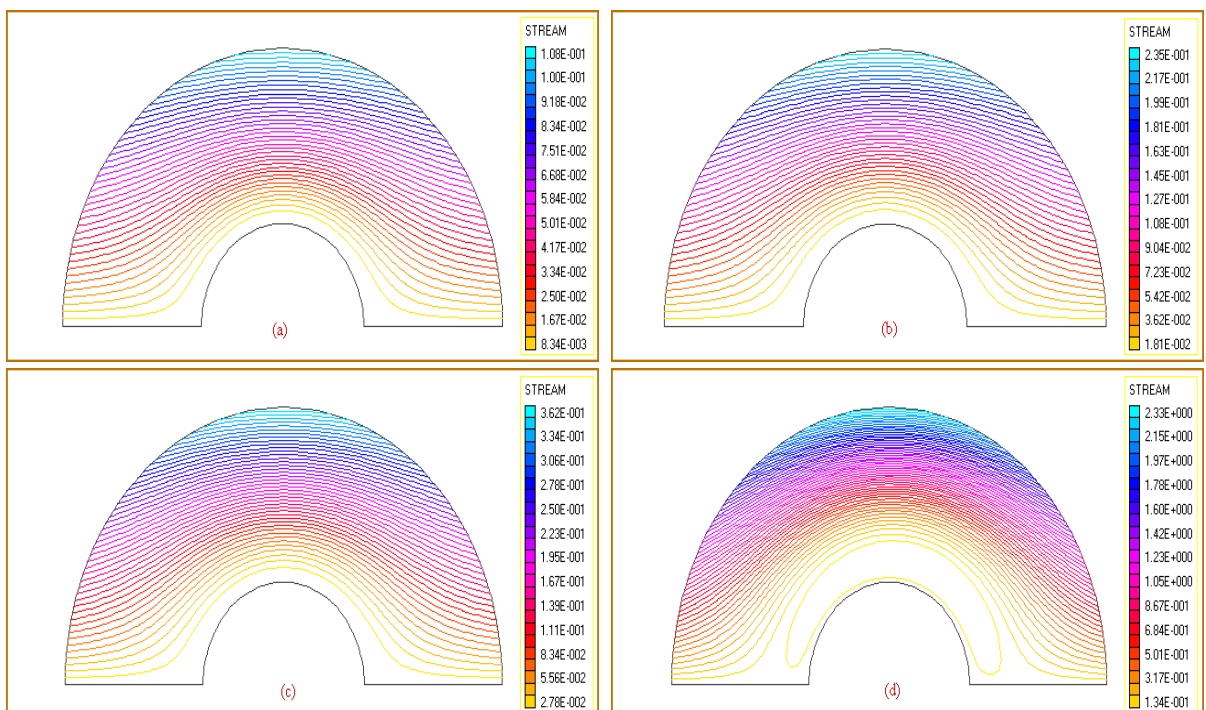

**Fig. 3. Streamlines for R=1, α=0 obtained by adopted numerical scheme at time levels (a) t=0.046 (b) t=0.1 (c) t=0.154 (d) t=0.993.**

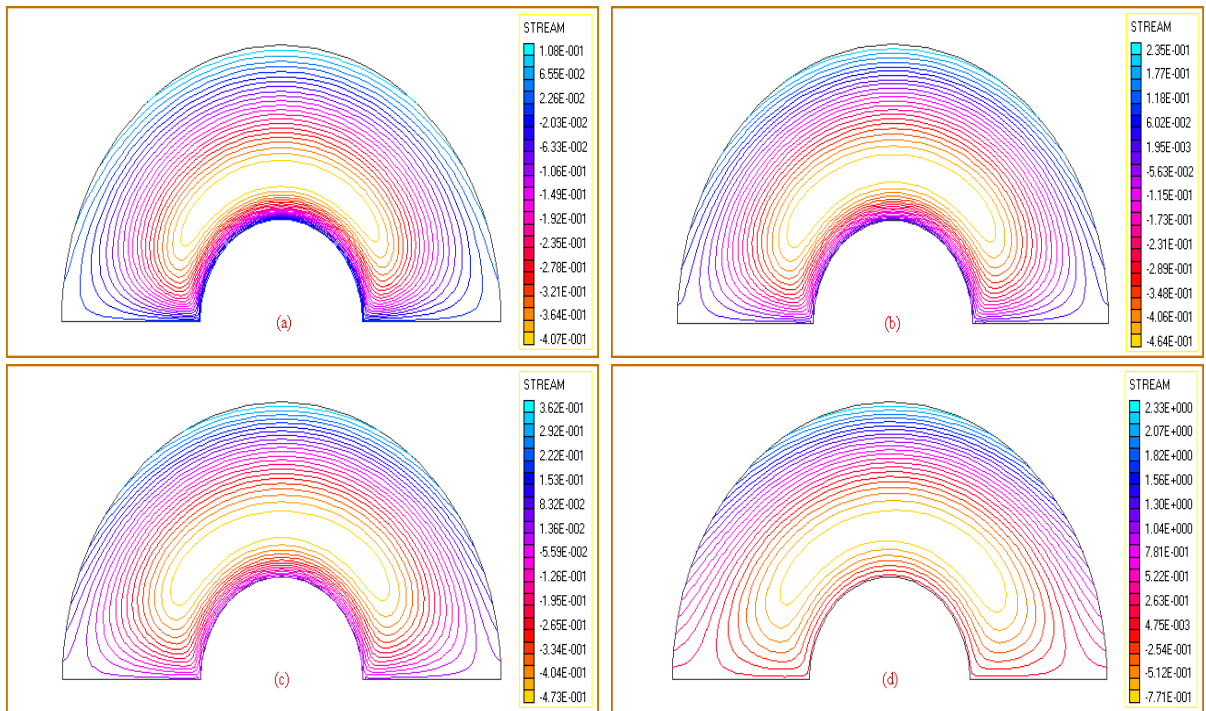

**Fig. 4. Streamlines for R=1, α=2.5 obtained by adopted numerical scheme at time levels (a) t=0.046 (b) t=0.1 (c) t=0.154 (d) t=0.993.**

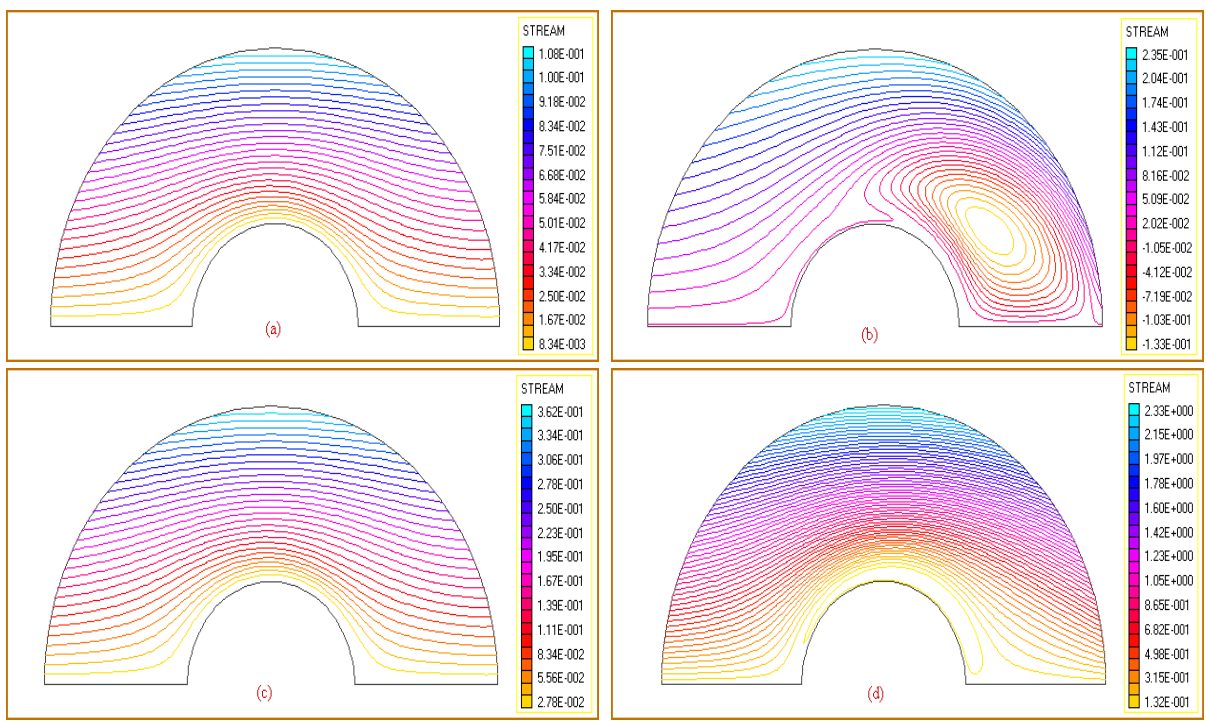

**Fig. 5. Streamlines for R=100, α=2.5 obtained by adopted numerical scheme at time levels (a) t=0.046 (b) t=0.1 (c) t=0.154 (d) t=0.993.**

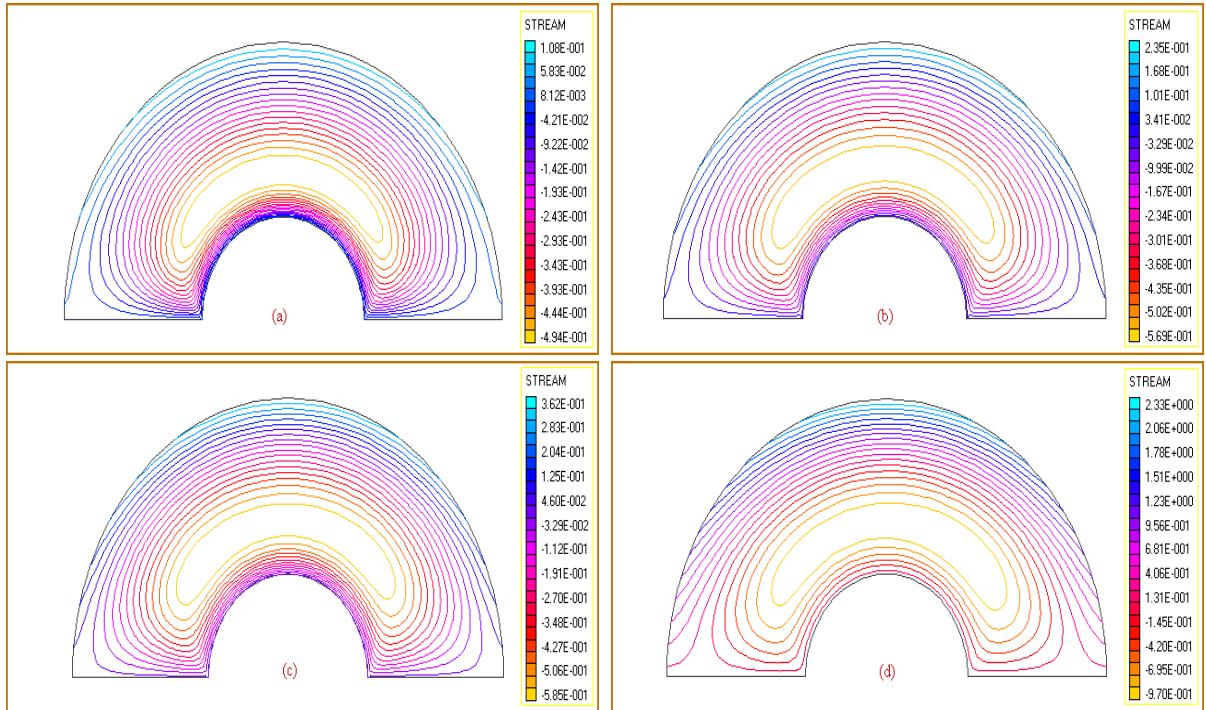

**Fig. 6. Streamlines for R=1, α=3.0 obtained by adopted numerical scheme at time levels (a)t=0.046 (b) t=0.1 (c) t=0.154 (d) t=0.993.**

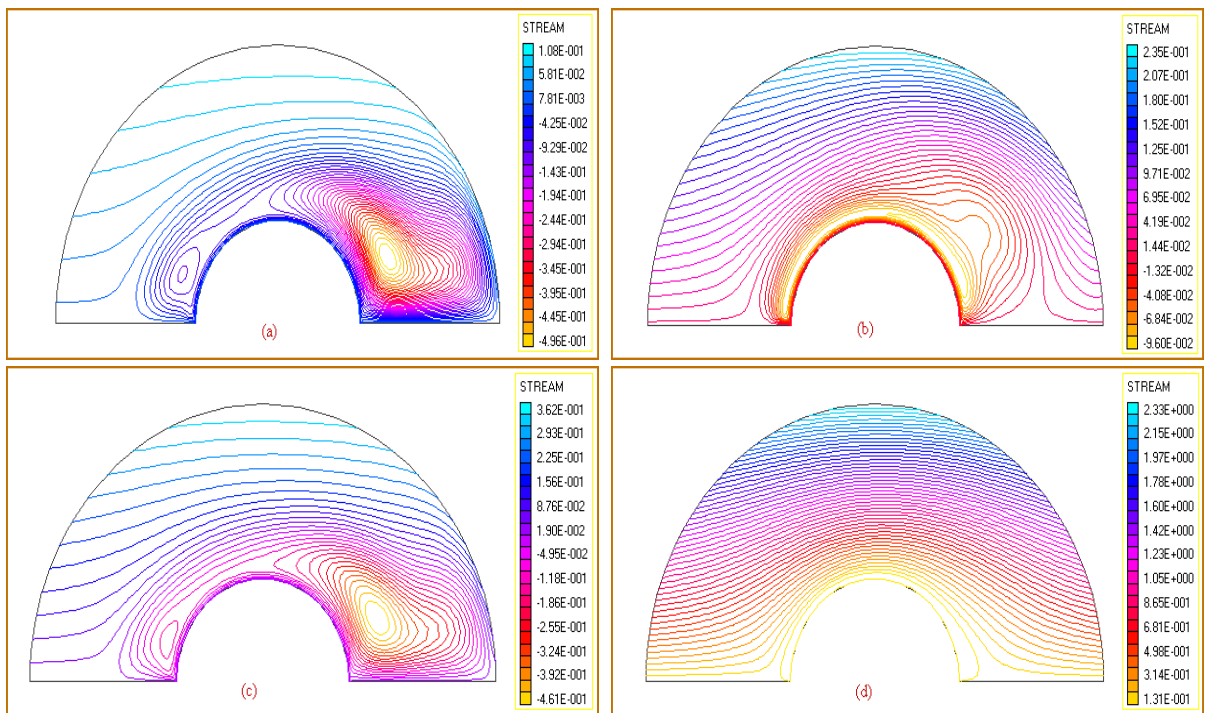

**Fig. 7. Streamlines for R=100, α=3.0 obtained by adopted numerical scheme at time levels (a)t=0.046 (b) t=0.1 (c) t=0.154 (d) t=0.993.**

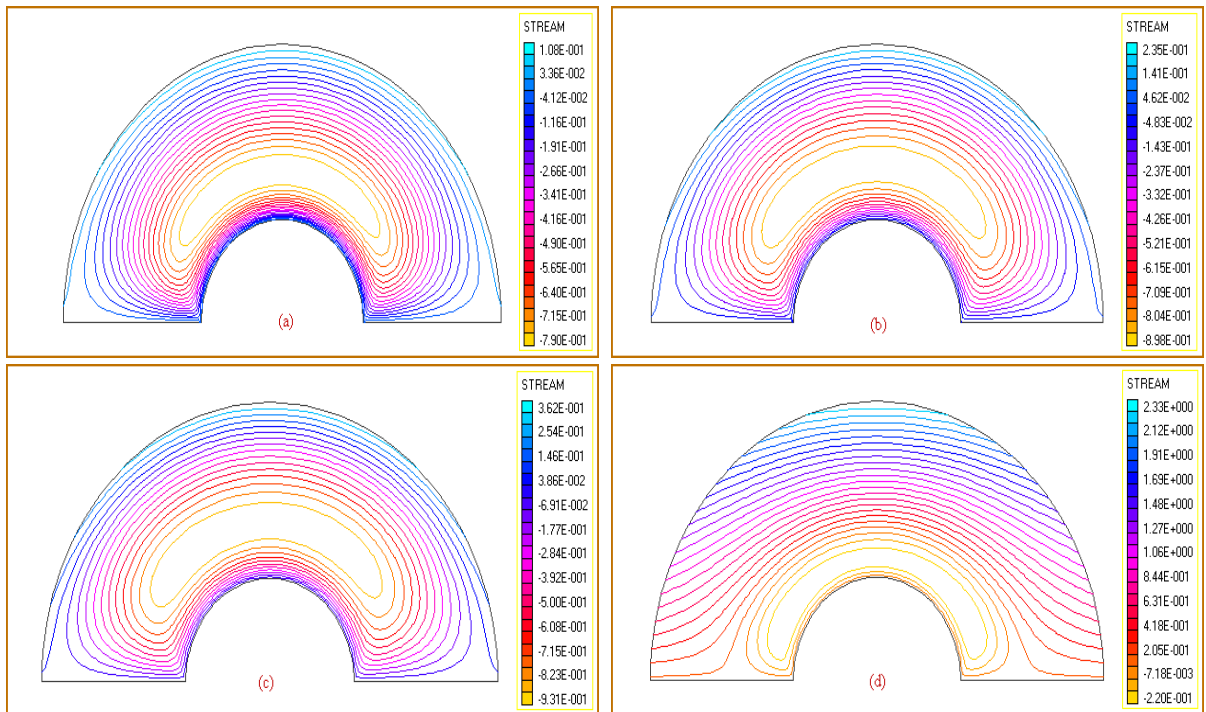

**Fig. 8.** Streamlines for R=1, α=4.5 obtained by adopted numerical scheme at time levels (a) t=0.046 (b) t=0.1 (c) t=0.154 (d) t=0.993.

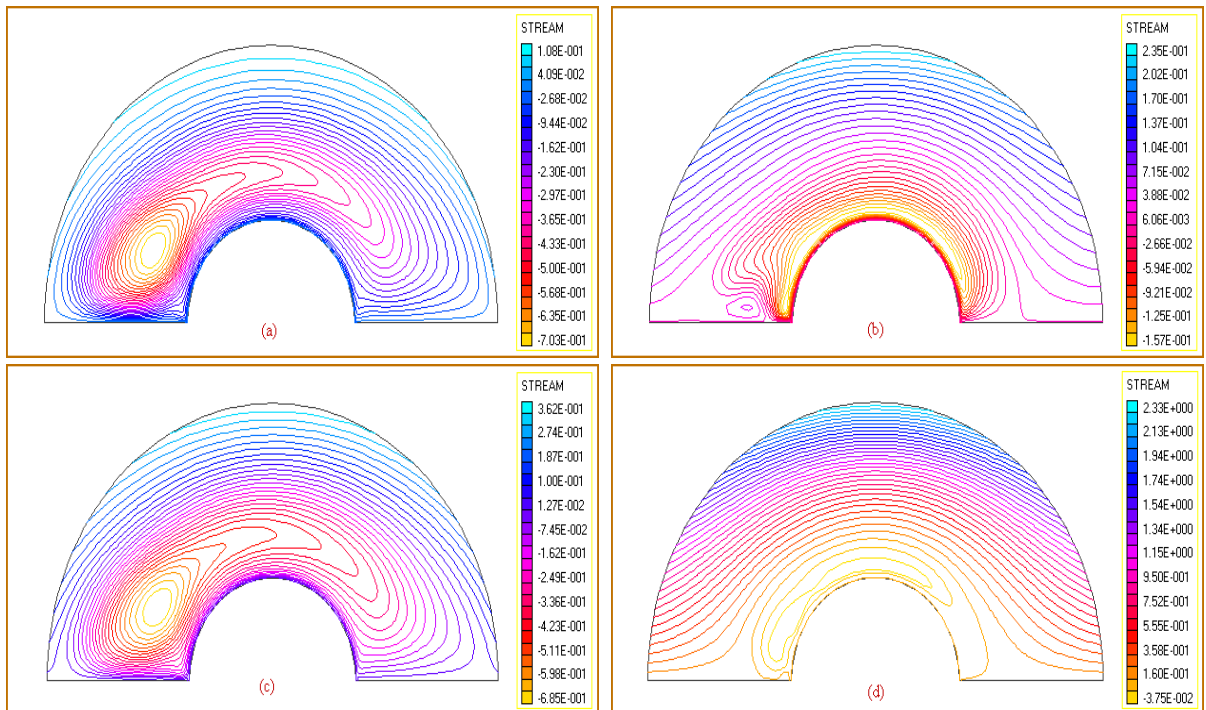

**Fig. 9.** Streamlines for R=100, α=4.5 obtained by adopted numerical scheme at time levels (a) t=0.046 (b) t=0.1 (c) t=0.154 (d) t=0.993.

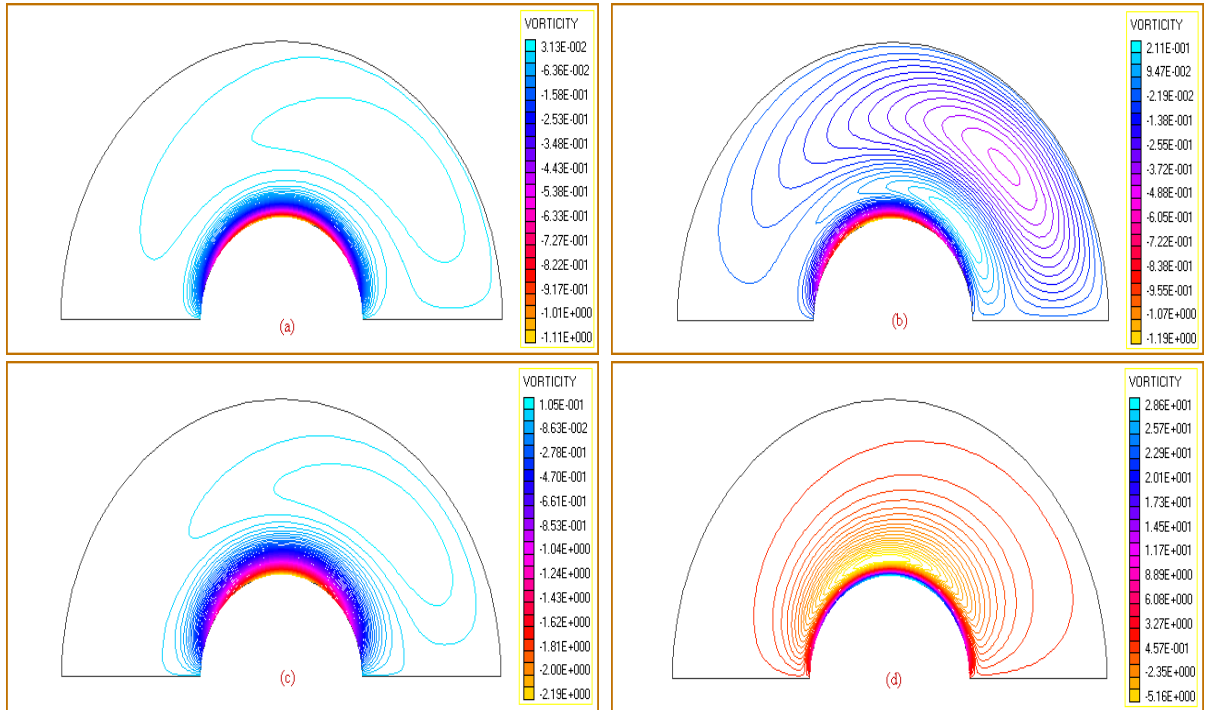

**Fig. 10. Equi-vorticity lines for R=10, α=0 at time levels (a) t=0.046 (b) t=0.1 (c) t=0.154 (d)t=0.993.**

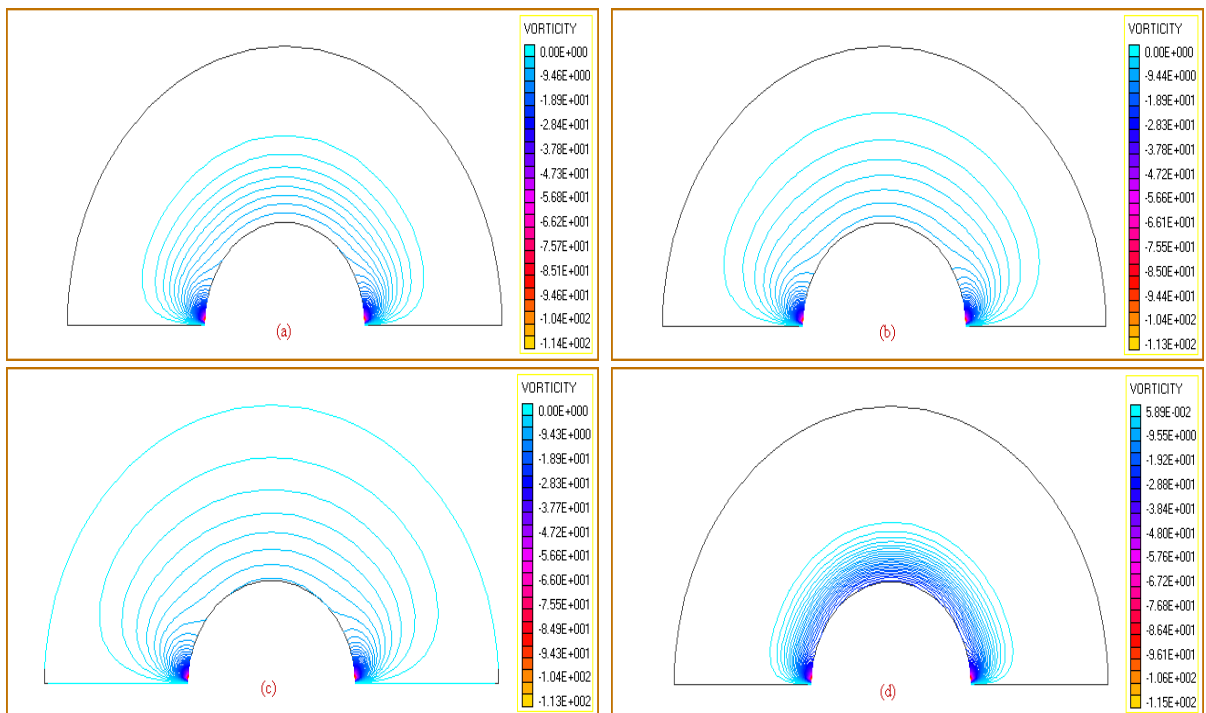

**Fig. 11. Equi-vorticity lines for R=10, α=4.5 at time levels (a) t=0.046 (b) t=0.1 (c) t=0.154 (d)t=0.993.**

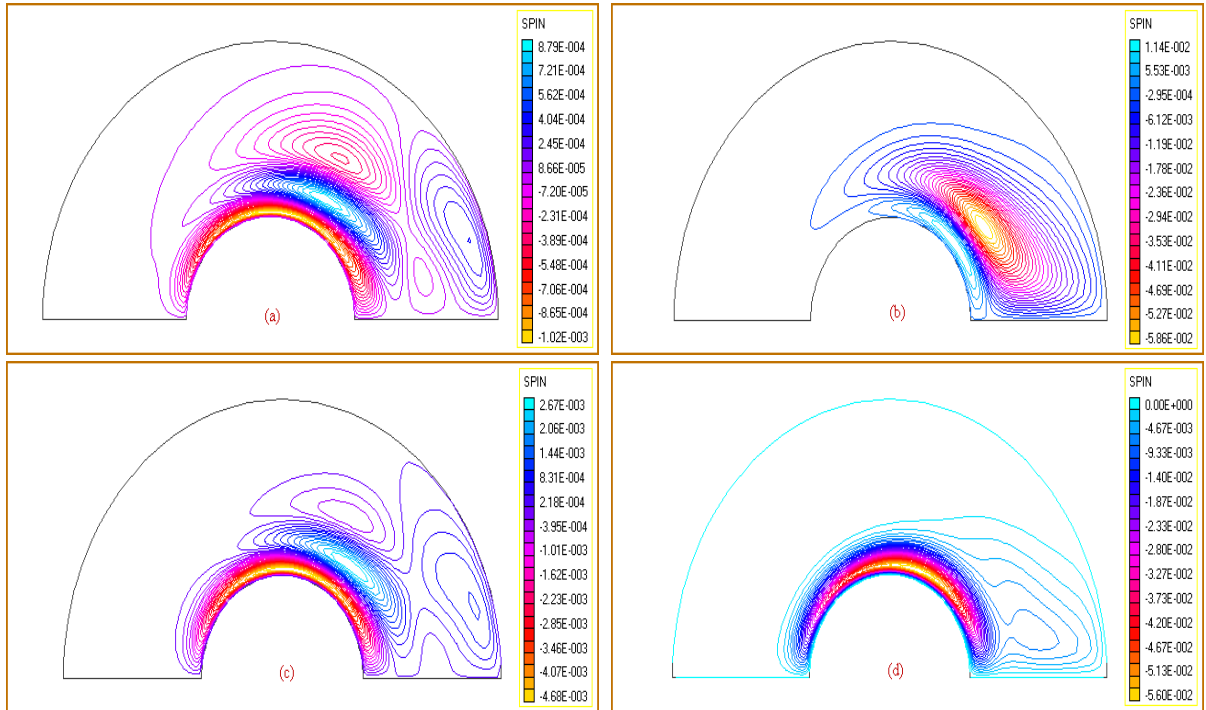

**Fig.12. Equi-spin lines for R=100, α=0 at time levels (a) t=0.046 (b) t=0.1 (c) t=0.154 (d) t=0.993.**

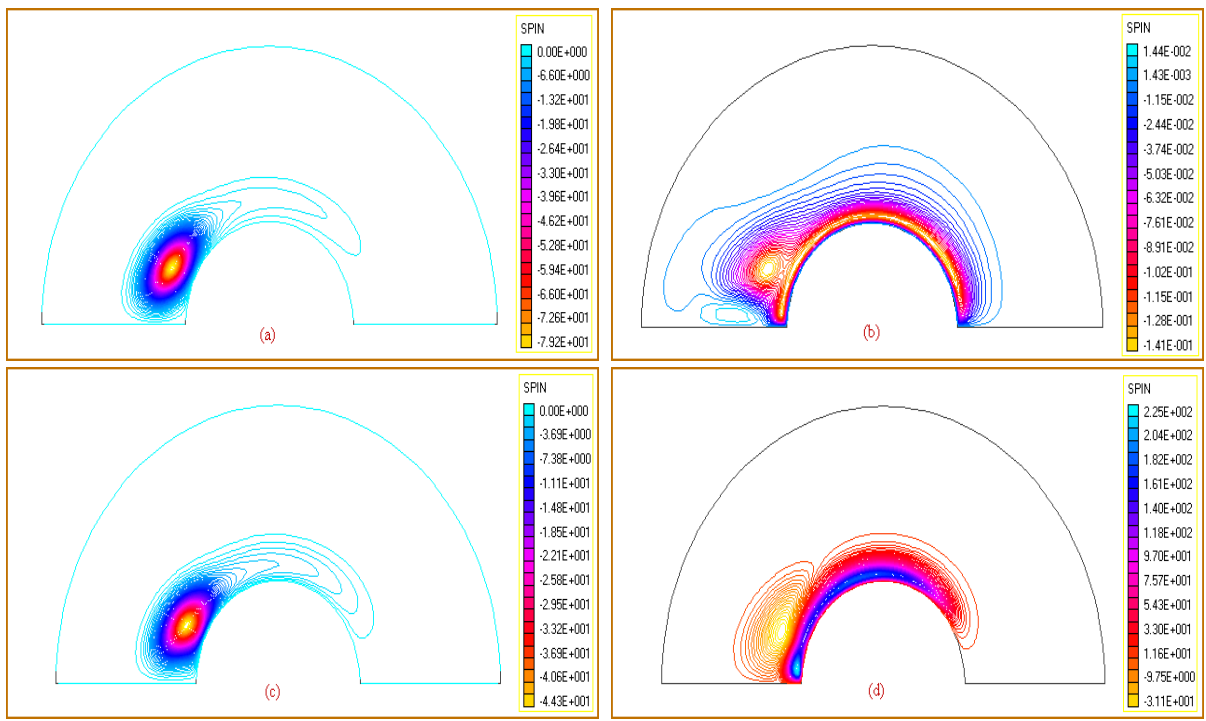

**Fig. 13. Equi-spin lines for R=100, α=4.5 at time levels (a)t=0.046 (b) t=0.1 (c) t=0.154 (d)t=0.993.**

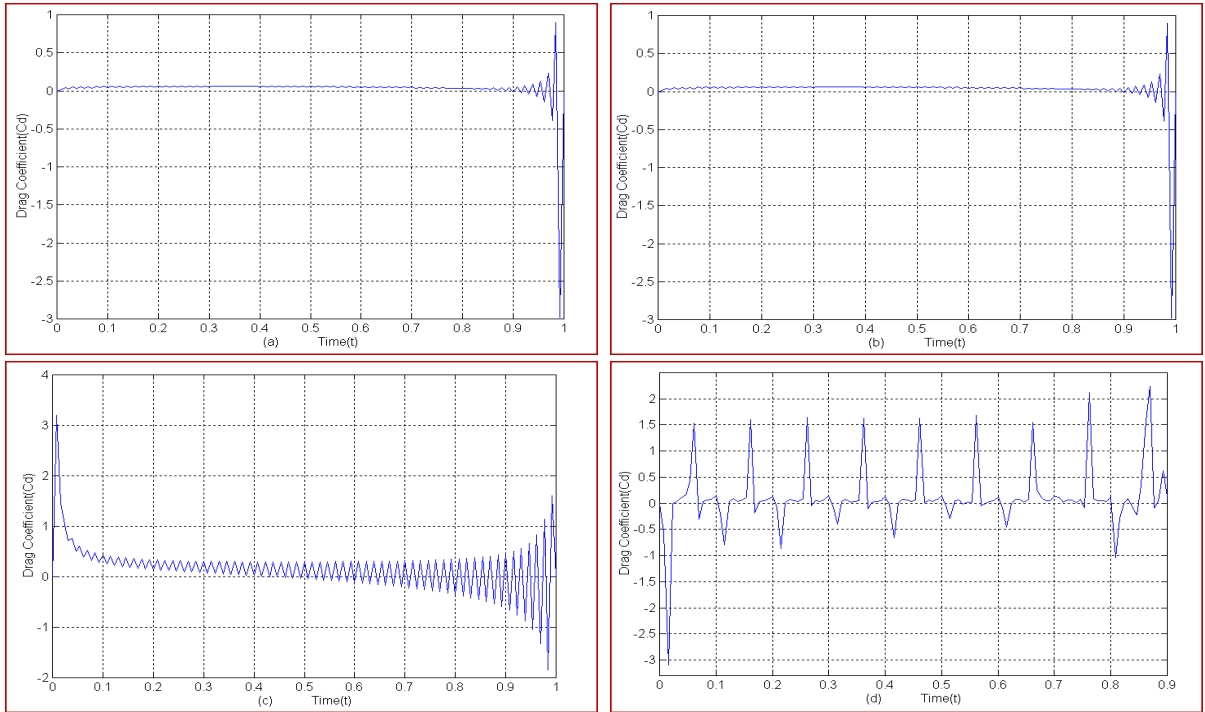

**Fig. 14. The variation of drag over the surface of cylinder for R=100 when (a) α=0 (b) α=2. 5 (c)α=3 (d) α=4. 5.**

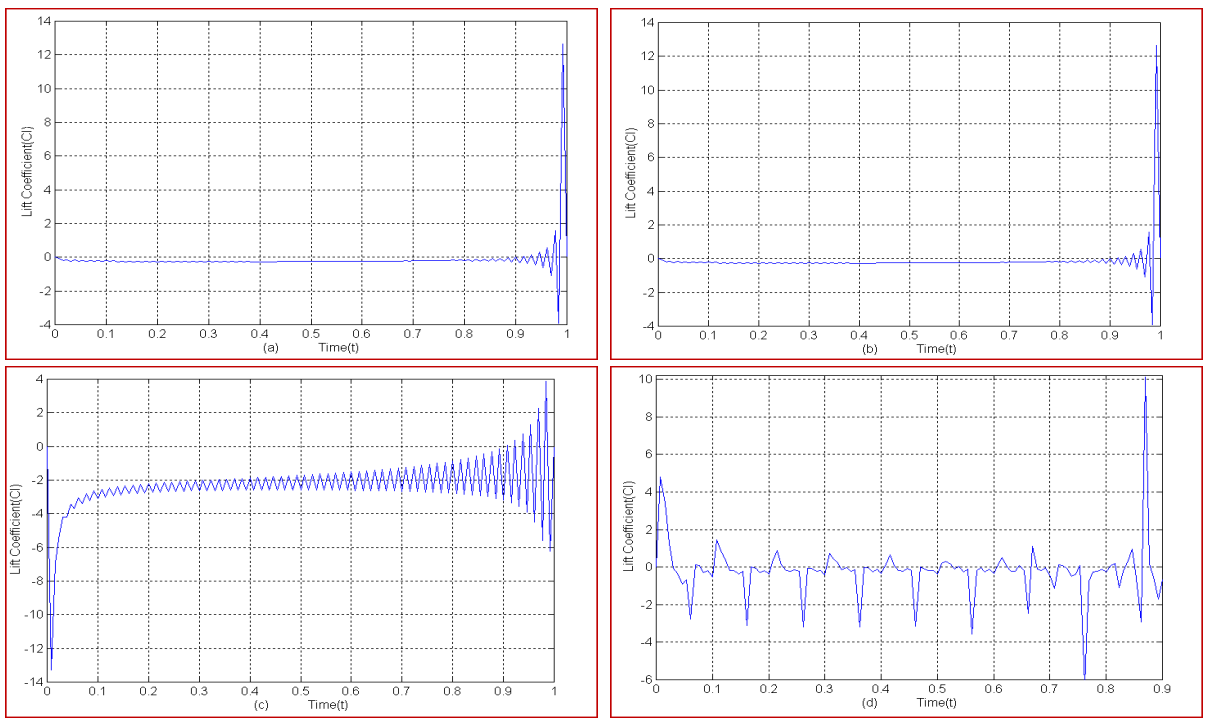

**Fig. 15. The variation of lift over the surface of cylinder for R=100 when (a) α=0 (b) α=2. 5 (c)α=3 (d) α=4. 5.**

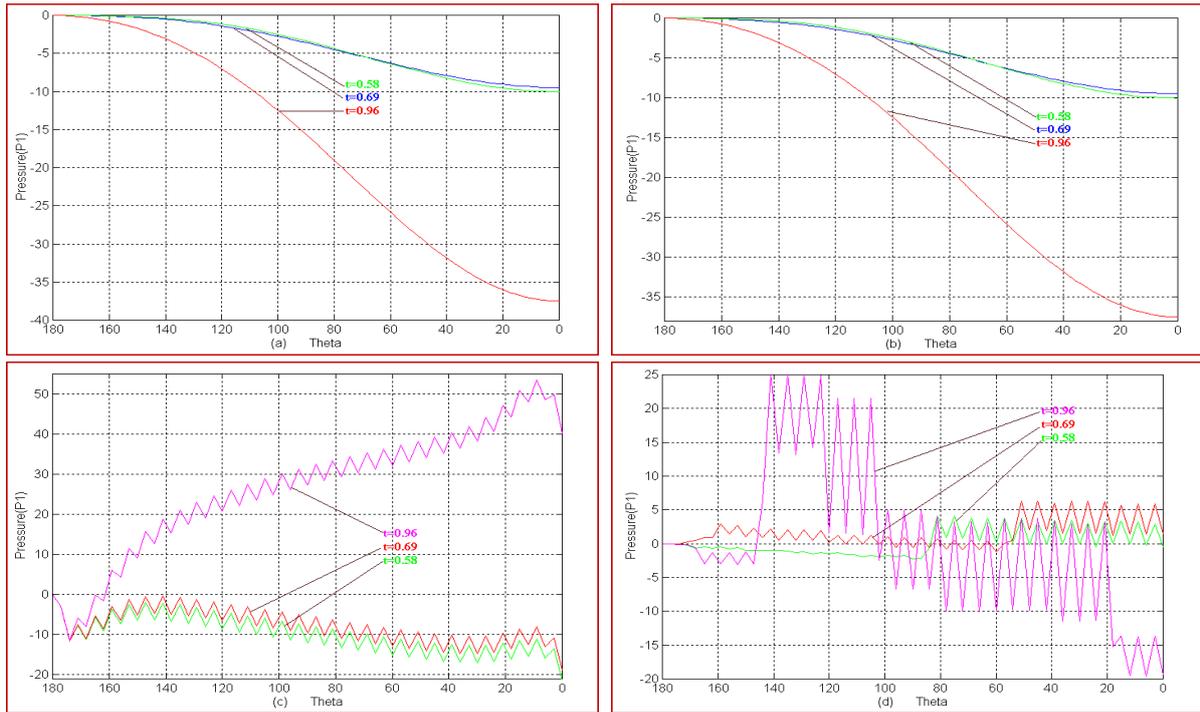

**Fig. 16. The variation of pressure over the surface of cylinder for R=100 when (a) α=0 (b) α=2. 5 (c) α=3 (d) α=4. 5.**

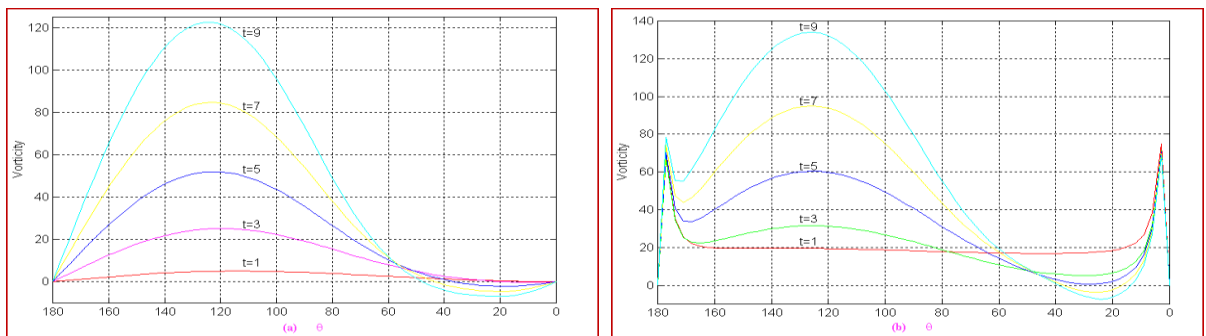

**Fig. 17. The variation of vorticity over the surface of cylinder for smaller time for R=10 when (a) α=0 (b) α=2. 5 (c) α=3 (d) α=4. 5.**

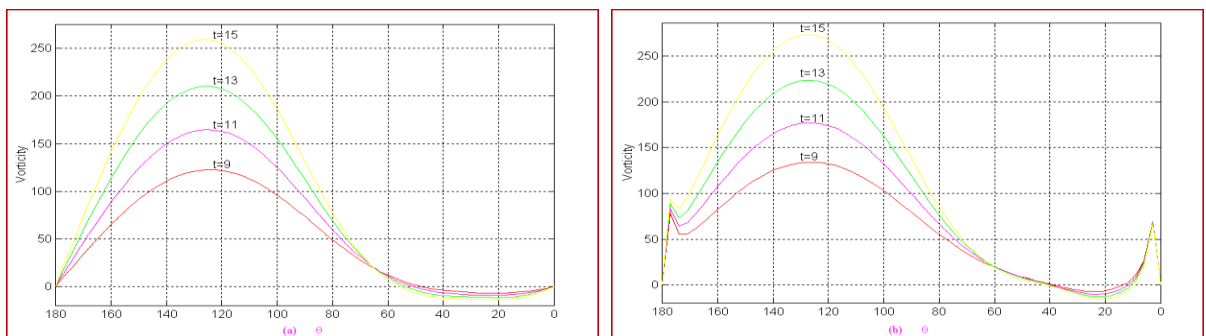

**Fig. 18. The variation of vorticity over the surface of cylinder for larger time for R=10 when (a) α=0 (b) α=2. 5 (c) α=3 (d) α=4. 5.**

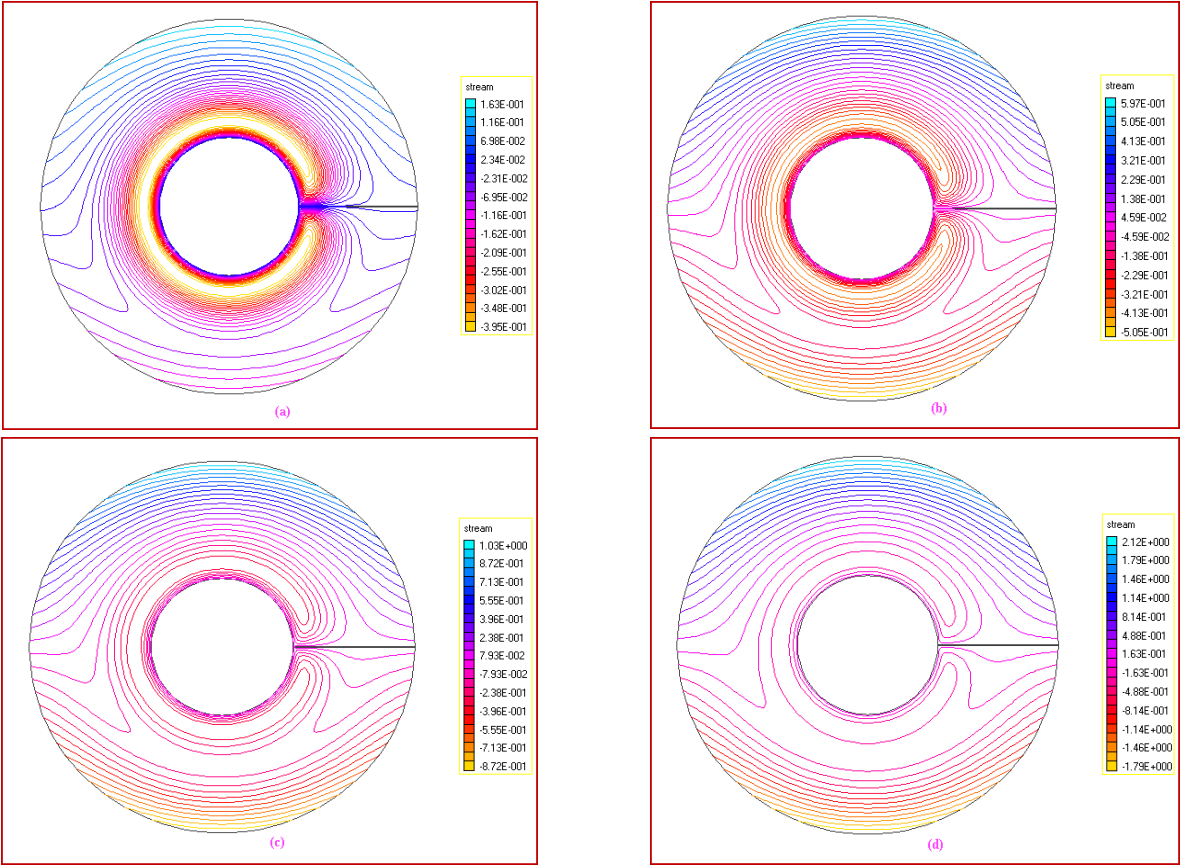

**Fig. 19. Streamlines for R=1, α=4.5 for case (b) at time levels (a)t=0.046(b)t=0.1 (c)t=0.154(d)t=0.993.**

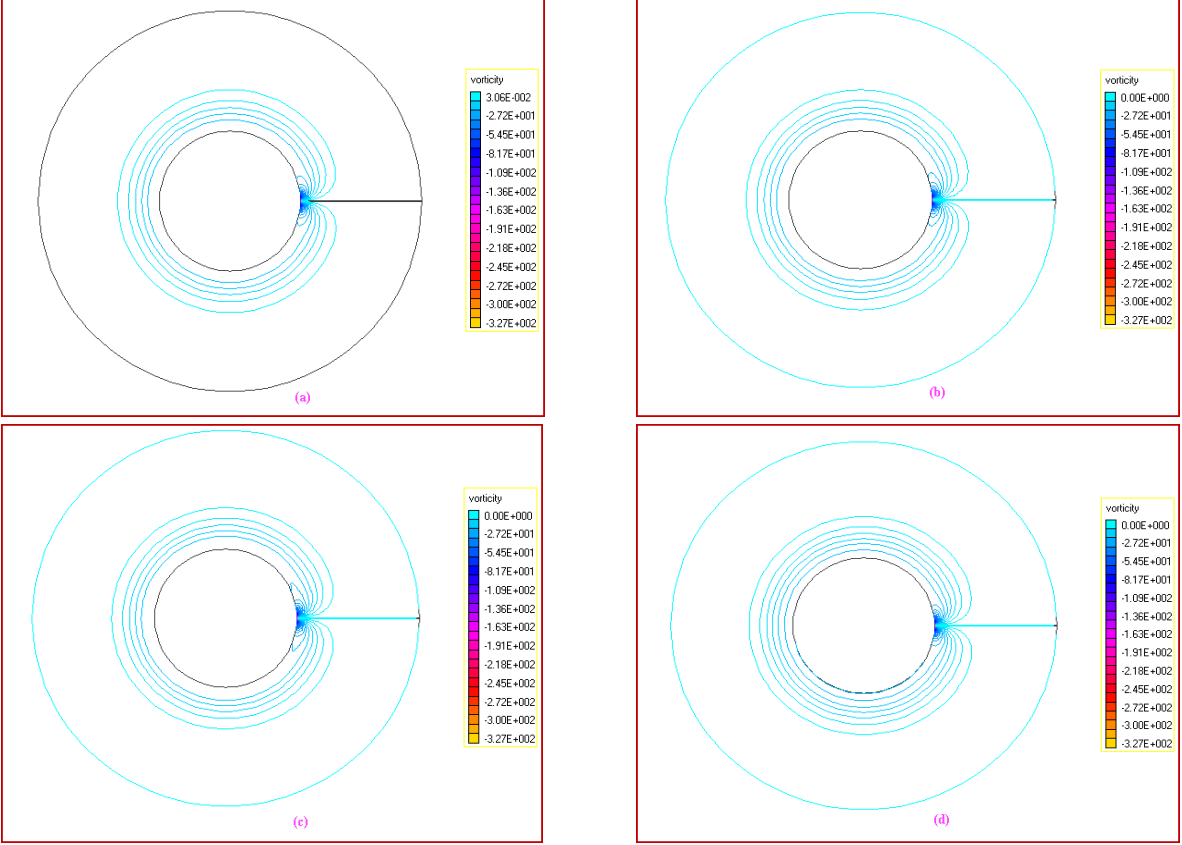

**Fig. 20. Equivorticitylines for R=1, α=4.5 for case b at time levels (a) t=0.046(b) t=0.1 (c) t=0.154 (d)t=0.993.**

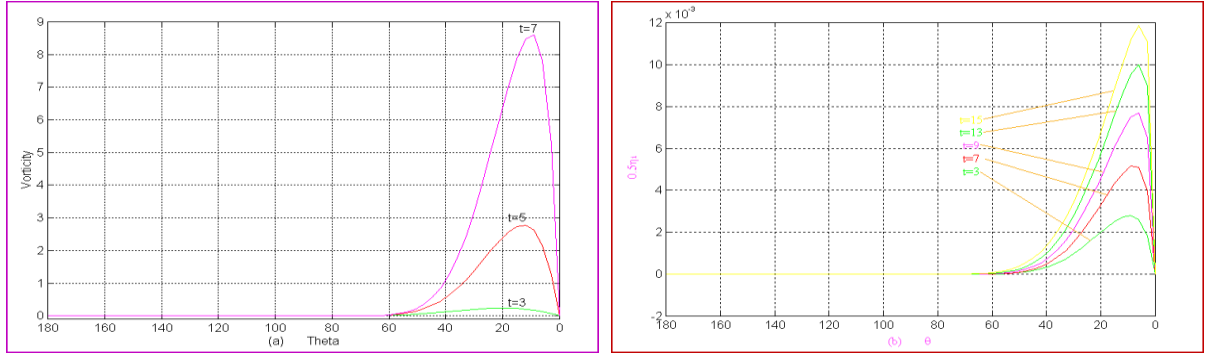

**Fig. 21.** The variation of (a) vorticity (b) spin, on outer boundary of the domain for $\alpha=0$, $R=10$.

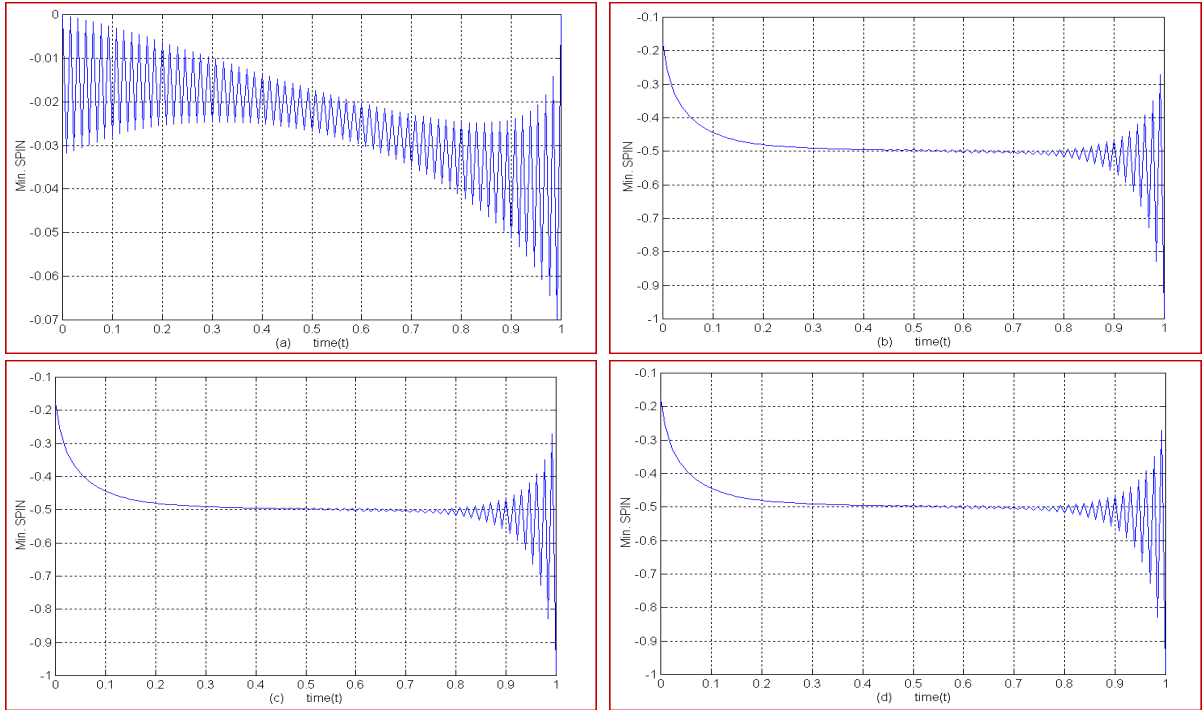

**Fig. 22.** The variation of $\eta_{min}$ for $R=10$ when (a) $\alpha=0$ (b) $\alpha=2.5$ (c) $\alpha=3$ (d) $\alpha=4.5$.

| t | $\alpha=0.0$ | | $\alpha=2.5$ | | $\alpha=3.0$ | | $\alpha=4.5$ | |
|---|---|---|---|---|---|---|---|---|
| | $\psi_{max}$ | | $\psi_{max}$ | | $\psi_{max}$ | | $\psi_{max}$ | |
| | By ABTFSM | By RKSFDM | By ABTFSM | By RKSFDM | By ABTFSM | By RKSFDM | By ABTFSM | By RKSFDM |
| 0.123 | 0.28917 | 0.28928 | 0.288930 | 0.28928 | 0.28800 | 0.28928 | 0.28860 | 0.28928 |
| 0.469 | 1.10280 | 1.10288 | 1.109980 | 1.10288 | 1.118992 | 1.10288 | 1.11931 | 1.10288 |
| 0.692 | 1.62723 | 1.62720 | 1.619882 | 1.62720 | 1.613988 | 1.62720 | 1.59997 | 1.62720 |
| 0.877 | 2.06113 | 2.06112 | 2.058777 | 2.06112 | 2.053331 | 2.06112 | 2.04572 | 2.06112 |

**Table 2.** The variation of $\psi_{max}$ with t and $\alpha$ by two methods.

| t | $\alpha=0.0$ | | $\alpha=2.5$ | | $\alpha=3.0$ | | $\alpha=4.5$ | |
|---|---|---|---|---|---|---|---|---|
| | $E_{min}$ | | $E_{min}$ | | $E_{min}$ | | $E_{min}$ | |
| | Newtonian fluid | Micropolar fluid | Newtonian fluid | Micropolar fluid | Newtonian fluid | Micropolar fluid | Newtonian fluid | Micropolar fluid |
| 0.123 | -2.07605 | -2.09314 | -72.5635 | -122.679 | -88.0638 | -122.679 | -136.701 | -122.679 |
| 0.469 | -3.52845 | -3.5767 | -72.1753 | -122.638 | -87.6392 | -122.638 | -136.065 | -122.638 |
| 0.692 | -6.84184 | -6.79654 | -71.7779 | -122.658 | -87.2022 | -122.658 | -135.952 | -122.658 |
| 0.877 | -9.81583 | -9.80686 | -71.1902 | -122.554 | 86.5689 | -122.554 | -135.372 | -122.554 |

**Table 3.** The variation of $E_{min}$ with t and $\alpha$ for two fluids.